\documentclass[aps,prd,showpacs,eqsecnum,twocolumn,nofootinbib,superscriptaddress]{revtex4-1}
\usepackage{amsmath,amssymb,graphicx,color,ulem,multirow}
\begin{document}

\title{Gravitational Collapse of Rotating Supermassive Stars including Nuclear Burning Effects}

\author{Haruki Uchida}
\affiliation{Yukawa Institute for Theoretical Physics, 
Kyoto University, Kyoto, 606-8502, Japan~} 

\author{Masaru Shibata} 
\affiliation{Center for Gravitational Physics, 
	Yukawa Institute for Theoretical Physics, Kyoto University, Kyoto, 
	606-8502, Japan~}
\author{Takashi Yoshida}
\affiliation{Department of Astronomy, Graduate School of Science,
	the University of Tokyo, Tokyo, 113-0033, Japan~}
\date{\today}

\author{Yuichiro Sekiguchi} 
\affiliation{Department of Physics, Toho
	University, Funabashi, Chiba 274-8510, Japan}
\author{Hideyuki Umeda}
\affiliation{Department of Astronomy, Graduate School of Science,
	the University of Tokyo, Tokyo, 113-0033, Japan~} 

\newcommand{\beq}{\begin{equation}}
\newcommand{\eeq}{\end{equation}}
\newcommand{\beqn}{\begin{eqnarray}}
\newcommand{\eeqn}{\end{eqnarray}}
\newcommand{\pa}{\partial}
\newcommand{\vp}{\varphi}
\newcommand{\varep}{\varepsilon}
\newcommand{\ep}{\epsilon}
\newcommand{\comp}{(M/R)_\infty}
\newcommand{\apjs}{ApJS}
\newcommand{\apjl}{ApJL}
\newcommand{\aap}{A{\&}A}
\newcommand{\aaps}{A{\&}AS}
\newcommand{\mnras}{MNRAS}
\begin{abstract}
Supermassive stars~(SMSs) of mass $\gtrsim 10^5 M_\odot$ are candidates for seeds of supermassive black holes found in the center of  many massive galaxies. We simulate the gravitational collapse of a rigidly rotating SMS core including nuclear burning effects in axisymmetric numerical-relativity simulation. We find that for realistic initial conditions, the nuclear burning does not play an important role. After the collapse, a torus surrounding a rotating black hole is formed and a fraction of the torus material is ejected.  We quantitatively study the relation between the properties of these objects and rotation. We find that if a SMS core is sufficiently rapidly rotating, the torus and outflow mass have approximately $6\%$ and $1\%$ of the initial mass, respectively. The typical average velocity and the total kinetic energy of the outflow are $0.2~c$ and $10^{54-56}$ erg where $c$ is the speed of light. Finally, we briefly discuss the possibility for observing the torus and outflow. 
\end{abstract}

\pacs{draft version}

\maketitle


\section{Introduction}
\label{intro}
Recent observations have revealed that there are many supermassive black holes~(SMBHs) in the center of massive galaxies. 
However, the formation process of SMBHs still remains unsolved. 
One possible scenario is the so-called direct-collapse scenario~\cite{2003ApJ...596...34B}. 
In this scenario, one supposes that a supermassive star (SMS) with mass $\gtrsim 10^5M_\odot$ is formed in a very hot primodial gas cloud with its virial temperature $\gtrsim 10^4$K, 
and it subsequently forms a high-mass seed black hole through gravitational collapse. 
We note that for such high-temperature environment, a mass-accretion rate to the growing SMS with $>0.1M_\odot/{\rm yrs}$ is possible \cite{2014MNRAS.445L.109I}.

During the gas accretion, the temporal mass accretion rate, $\dot{M}_{\rm BH}$, is naively bounded by the Eddington rate such that
\begin{equation}
\dot{M}_{\rm BH} = \frac{1-\xi}{\xi} \frac{4\pi G m_{\rm p}}{c\sigma_{T}} M_{\rm BH},
\label{i-1}
\end{equation}
where $G,c,~\xi,~m_{\rm p},~\sigma_T$, and $M_{\rm BH}$ are the gravitational constant, the speed of light, the energy conversion rate by accretion ($<1$), the mass of proton, the Thomson scattering cross section, and the temporal mass of the black hole, respectively. Then we can estimate the growth time by solving Eq.~(\ref{i-1}) and get
\begin{equation}
t_{\rm growth} \approx 0.12 \times {\rm log}_{10} \left ( \frac{M_{\rm BH}}{M_{\rm seed}}\right ) {\rm Gyr},
\label{i-2}
\end{equation} 
where $M_{\rm seed}$ is the mass of the seed black hole and we take $\xi =0.1$.
For a SMBH with mass $10^7M_\odot$, which is the typical mass of SMBHs in local spiral galaxies~\cite{2014ApJ...789..124D},
and for $M_{\rm seed}=100M_\odot$, which is a typical value of first stars \cite{2015MNRAS.448..568H}, we get $t_{\rm growth} =0.58$ Gyr. 
On the other hand, in the direct-collapse scenario, inserting $M_{\rm seed}=10^5M_\odot$ and $M_{\rm BH} =10^7M_\odot$ to Eq.~(\ref{i-2}), we get $t_{\rm growth}=0.23$ Gyr.
Thus the condition for the mass accretion rate to form SMBHs is relaxed.  
We note that this scenario is also thought to be one possible scenario which can form SMBHs in the early universe at redshift $z > 6$~(e.g.,~Refs.~\cite{2015Natur.518..512W, 2011Natur.474..616M}).

Recent researches for SMS formation in spherical
symmetry~(e.g.,~Refs.~\cite{2013ApJ...778..178H,2016ApJ...830L..34U}) have proposed that 
a SMS with mass $\gtrsim 2\times 10^5M_\odot$ could be formed if the mass-accretion rate reaches $0.1M_\odot /$yrs (i.e., the temperature of a primodial gas cloud becomes $\gtrsim 10^4~{\rm K}$) and lasts for more than the period of nuclear-burning phases~($\approx 2\times 10^6$ yrs~\cite{1984ApJ...280..825B}). 
To achieve such high virial temperature, there should not exist molecular hydrogen (${\rm H}_2$) in a primordial gas cloud. In the absence of ${\rm H}_2$, the gas cloud could reach this virial temperature because atomic hydrogen cooling could achieve only about $10^4$K~\cite{2001ApJ...546..635O}. 
There are several routes to destroy ${\rm H}_2$ molecules such as photodissociation by Lyman-Werner radiation from nearby local star formation regions~\cite{2001ApJ...546..635O, 2016ApJ...832..134C} or collisional dissociation in the cold accretion flows in the forming first galaxies~\cite{2012MNRAS.422.2539I}. 

Latest numerical simulations suggest that SMSs are rotating because the environments surrounding each protostar of SMSs are 
not spherically symmetric~(e.g.,~Refs.~\cite{2013MNRAS.430..588L, 2014MNRAS.439.1160R, 2015MNRAS.446.2380B}). 
Also SMS cores seem to be rigidly rotating because convection is strongly enhanced if they are in nuclear-burning phases~\cite{1984ApJ...280..825B,1994ApJ...432...52L,2016ApJ...830L..34U}. 

If SMSs have sufficiently large mass, they may collapse due to the so-called general-relativistic radial instability~(e.g.,~Ref.~\cite{1964ApJ...140..417C}). 
If a SMS core is rotating, the condition for a SMS core to become unstable to the gravitational collapse 
is greatly different from the non-rotation case because rotation strongly stabilizes a SMS core~(e.g.,~Refs.~\cite{1945ApJ...102..143L,1966ApJ...144..180F,1978trs..book.....T,1999ApJ...526..941B,0004-637X-818-2-157}). 
Our previous result suggests that SMS cores can be stable unless their mass exceeds about $6. 3\times 10^5 M_\odot$ in the hydrogen-burning phase and $2. 3 \times 10^5M_\odot$ in the helium-burning phase if they are rotating at mass-shedding limit~\cite{0004-637X-818-2-157}. These critical values are about $5$ times larger than those for non-rotating SMSs. 

SMSs have not been directly observed yet. 
However, in its presence, the gravitational collapse of SMS cores could be observed. 
Our previous study~\cite{PhysRevD.94.021501} proposed that if a SMS core is rotating, gravitational waves associated with the quasi-normal mode ringdown are emitted during the black-hole formation and if it occurs at the cosmological redshift less than $\approx 3$, the signal will be detectable by space laser interferometric detectors like LISA \cite{2017arXiv170200786A}. 
Other studies show that a collapsing SMS may be detectable as a gamma-ray burst or an ultra-luminous supernova if the formed black hole launches a relativistic jet during the collapse~\cite{2015ApJ...810...64M,2016ApJ...823...83M}. 

The primary purpose of this paper is to explore the effects of nuclear burning in the collapse of SMSs and for the remnants of the SMS collapse. 
For the effect of nuclear burning, there are several pioneering  studies that indicate that its effect may change the SMS collapse into an explosion like pair instability supernovae~\cite{1986ApJ...307..675F,2012ApJ...749...37M,2014ApJ...790..162C}. 
However, they focused only on a restricted class of the gravitational collapse of SMS cores. 
In Refs.~\cite{1986ApJ...307..675F,2012ApJ...749...37M}, the authors considered SMS cores composed of hydrogen and free from nuclear burning
 and discovered that a SMS core would explode by the ignition of the hydrogen during the collapse phase if its initial metallicity is larger than $O(10^{-3})$.
In Ref.~\cite{2014ApJ...790..162C}, they considered non-rotating SMSs and concluded that a SMS with mass close to $\approx 55500M_\odot$ would explode due to helium burning. 
In reality, the typical metallicity of the SMS core at hydrogen burning phase will be $O(10^{-9})$~\cite{1984ApJ...280..825B} and it is natural to consider that it would be rapidly rotating~\cite{2015MNRAS.446.2380B}. Moreover, no study has paid special attention to the evolution of the remnant formed after the SMS core collapse.

In this paper, we perform general-relativistic simulations of the gravitational collapse of rotating SMS cores from plausibly realistic initial conditions including the effects of nuclear burning and rotation. We will show that for the initial conditions we employed, a black hole is formed irrespective of the presence of the nuclear burning effect. In addition, a torus surrounding the black hole is formed. We will also show that for the evolution of the torus, the nuclear burning effect does not play an important role. 

After the black-hole formation, a fraction of the torus material is ejected as an outflow. 
We will describe the formation process of the outflow (see Ref.~\cite{PhysRevD.76.084017} as a pioneering study).   
We will show that if the initial SMS core is sufficiently rapidly rotating, the typical total kinetic energy and speed of the outflow are $10^{54-56}$ erg and $0.2c$, respectively. 
 
The paper is organized as follows. In Sec.~\ref{setup}, we describe the setup of our numerical simulation. In Sec.~\ref{result}, we describe the overview of the collapse showing our results of numerical simulations and discuss the effects of nuclear burning. 
We also study the dependence of the mass of the torus surrounding a black hole formed after the collapse on the rotation and adiabatic constant. In Sec.~\ref{outflow}, we describe the formation 
process of the outflow and explore its properties. In Sec.~\ref{discussion}, we discuss the possibility for observing of the torus by the gravitational-wave observation. Section~\ref{conclusion} is devoted to the conclusion. 

\section{Numerical setup}
\label{setup}
\subsection{Calculation of gravitational field}
For solving Einstein's evolution equations, we use the same method as in Ref.~\cite{PhysRevD.94.021501}. 
We employ the original version of BSSN~(Baumgarte-Shapiro-Shibata-Nakamura) formalism with a puncture gauge~\cite{1995PhRvD..52.5428S,1999PhRvD..59b4007B,2006PhRvL..96k1101C,PhysRevLett.96.111102}.  
{In the $3+1$ formulation, the metric is defined by the form 
\begin{equation}
ds^2 = -\alpha^2 c^2dt^2  + \gamma_{ij} (dx^i + \beta^i c dt)(dx^j + \beta^j c dt),
\label{metric}
\end{equation}
where $\alpha,~\beta^i$, and $\gamma_{ij}$ are the lapse function, the shift vector, and the induced metric on three-dimensional (3D) spatial hypersurfaces, respectively. 
We also define the extrinsic curvature by
\begin{equation}
K_{ij}\equiv -\gamma_i^{~\alpha} \gamma_j^{~\beta} \nabla_\alpha n_\beta,
\label{ext}
\end{equation}
where $n^{\mu}$ is a timelike unit-normal vector orthogonal to the 3D hypersurface. 
In the BSSN formalism, we evolve $\rho_{\rm g} \equiv ({\rm det}\gamma_{ij})^{-1/6}$,~ $\tilde \gamma_{\ij}\equiv \rho_{\rm g}^2 \gamma_{ij}$, 
$\tilde{A}_{ij} \equiv \rho_{\rm g}^2 (K_{\ij}-\gamma_{ij}K^k_k/3)$, $K^k_k$, and $F_i \equiv \delta^{jk}\partial_j \tilde{\gamma}_{ik}$.}
We use the standard 4th-order finite differencing scheme to solve the gravitational-field equations~(see chapter 3 of \cite{2016nure.book.....S} for a review). 

A previous work indicates that if a SMS core is rigidly rotating, there would be essentially no nonaxisymmetric deformation during the collapse~\cite{PhysRevD.71.024014}. 
Hence we assume the axial symmetry and use a 4th-order cartoon method to impose this condition to the gravitational field~\cite{2001IJMPD..10..273A,PhysRevD.67.024033}. 
We neglect viscosity because the timescale of the gravitational collapse is much shorter than the viscous timescale. 
We only consider the collapse of SMS cores because the density of their envelope is very low, and hence, they are unlikely to affect the collapse dynamics. 

We perform numerical simulations in cylindrical coordinates $(X, Z)$, and a nonuniform grid is used for $X$ and $Z$ in the following manner. 
We define the grid spacing at the center by $\Delta X_0 \equiv X_1 -X_0$.　Here $X_0 =0$ and $X_{\rm i}$ is the location of $i$-th grid. 
We use different manners of grid spacing inside and outside of a grid $X_{\rm in}$. 
For $X_i<X_{\rm in}$, $\Delta X_i \equiv X_i -X_{i-1} = \Delta X_0({\rm const})$, and for $X \geq X_{\rm in}$, $\Delta X_i = \eta \Delta X_ {i-1}$,  where $\eta$ is a constant. $\eta$ determines the nonuniform degree of the grid spacing. 
We set $X_{\rm in} \sim R_{\rm M}$ where $R_{\rm M}$ is the gravitational radius defined by
\begin{equation}
R_{\rm M} \equiv \frac{GM_0}{c^2},
\end{equation}
and $M_0$ is the initial mass of the SMS core.

We set outer boundaries of the computational domain at $\approx 600R_{\rm M}$ along each axis. 
To calculate the propagation of the outflow, we expand the computational domain to $\approx 4800R_{\rm M}$ along each axis after the formation of the central black hole.
We set $\Delta X_0 \approx 0.037R_{\rm M}, \eta = 1. 018$ for the low-resolution case, $\Delta X_0 \approx 0.027R_{\rm M} , \eta = 1. 017$ for the middle-resolution case, and $\Delta X_0 \approx 0.023R_{\rm M} , \eta = 1. 014$ for the high-resolution case. 
We show that the numerical results have a good convergence property for our models in Sec.~\ref{result}.

\subsection{Equations of the fluid}
To self-consistently calculate the effect of nuclear burning, we follow the method employed in Ref.~\cite{2016MNRAS.456.1320T}.
We introduce two densites, i.e., the rest-mass density, $\rho_0$, and the baryon density, $\rho$, defined respectively by
\begin{eqnarray}
\rho_0 &=& \sum_{\rm i} m_{\rm i}n_{\rm i} \ ({\rm i=p,\alpha,C,e}), \label{a-1} \\
\rho &=& m_{\rm u} (n_{\rm p}+4n_{\rm \alpha}+12n_{\rm C})=m_{\rm u}n_{\rm B}, \label{a-2}
\end{eqnarray}
where $m_{\rm u}$ and $n_{\rm B}$ are the atomic mass unit and the baryon number density. We use $m_{\rm i}$ and $n_{\rm i}$ for the rest mass and number density of the i-th species. Subscripts p, $\alpha$, C, and e denote H, $^4$He, $^{12}$C, and electron, respectively. 
We also define the density of each nucleus by
\begin{equation}
\rho_{\rm i} \equiv m_{\rm u}  A_{\rm i} n_{\rm i} \ ({\rm i=p,\alpha,C}),
\label{a-2.1}
\end{equation}
where $A_{\rm i}$ is the mass number for each nucleus.
According to the electrical charge neutrality, $n_{\rm e}$ can be written as
\begin{equation} 
n_{\rm e}= n_{\rm p} + 2n_{\rm \alpha} + 6n_{\rm C}.
\label{s-3.1}
\end{equation}
Note that $\rho$ is proportional to the baryon number density, and thus, it does not change by nuclear burning. 
Hence it is convenient to define thermodynamic quantities in terms of $\rho$.

We assume a perfect fluid and the energy momentum tensor is written as
\begin{equation}
T_{\mu \nu} = \rho h u_\mu u_\nu + Pg_{\mu \nu},
\label{s-4}
\end{equation}
where
\begin{equation}
h \equiv \frac{\rho_0}{\rho}c^2 + \epsilon + \frac{P}{\rho},
\label{s-5}
\end{equation}
and $\epsilon, h, P$, and $u_\mu$ are the internal energy per baryon, the enthalpy per baryon, pressure, and four velocity, respectively.
We should be careful that $\epsilon$ and $h$ are not equivalent to the specific internal energy and specific enthalpy, respectively. 
We solve the conservation equations for the energy momentum tensor,
\begin{eqnarray}
\nabla_\mu T^{\mu \nu} = 0.
\label{s-8.1}
\end{eqnarray}

We shall mention the reason that we introduce $\rho$ and $\rho_0$ independently.
The difference between $\rho$ and $\rho_0$ is defined by
\begin{equation}
\delta \equiv \frac{\rho_0}{\rho} -1 =\Delta_{\rm p} Y_{\rm p} +\Delta_{\rm \alpha} Y_{\rm \alpha} +\Delta_{\rm C} Y_{\rm C},
\label{s-6}
\end{equation}
where
\begin{eqnarray}
\Delta_{\rm p} &=&\frac{1}{m_{\rm u}}(m_{\rm p} +m_{\rm e} -m_{\rm u}), \\
\Delta_{\rm \alpha} &=&\frac{4}{m_{\rm u}}\left (\frac{m_{\rm \alpha}}{4} +\frac{m_{\rm e}}{2} -m_{\rm u} \right ),  \\
\Delta_{\rm C} &=&\frac{12}{m_{\rm u}} \left (\frac{m_{\rm C}}{12} +\frac{m_{\rm e}}{2} -m_{\rm u} \right ), 
\end{eqnarray}
and $Y_{\rm i} \equiv n_{\rm i}/n_{\rm B}= \rho_{\rm i}/(\rho A_{\rm i})$. 

Now, we demonstrate how the rest-mass energy is converted to the internal energy via nuclear burning.
For simplicity, we shall use the fluid rest frame in the following analysis.
Then, the rest-mass energy density released via the nuclear burning can be written as
\begin{equation}
dE=-d\rho_0 c^2,
\label{s-6.1}
\end{equation}
where $d$ denotes the difference between before and after the nuclear burning. 
Using Eq.~(\ref{s-6}) and $d\rho=m_{\rm u}dn_{\rm B}=0$, Eq.~(\ref{s-6.1}) can be rewritten as 
\begin{equation}
dE = -\rho c^2d\delta.
\label{s-6.2}
\end{equation}
This equation denotes that the released rest-mass energy is proportional to the variation of $\delta$. 
The total energy density should be conserved, and hence,  
\begin{equation}
d(\rho_0 c^2 + \rho \epsilon)=0.
\label{s-6.3}
\end{equation}
Inserting Eqs.~(\ref{s-6.1}) and (\ref{s-6.2}) to Eq.~(\ref{s-6.3}), we get 
\begin{equation}
d(\rho \epsilon) =dE=-\rho c^2d\delta.
\label{s-6.4}
\end{equation}
Therefore the released rest-mass energy is autonomically converted to the increase of the internal energy via the decrease of $\delta$.
Actually, some fraction of the released rest-mass energy is emitted by neutrinos, and thus, this formalism slightly overestimates the 
increase of the internal energy.

If we consider the effect of nuclear burning, the continuity equations for each nucleus should be written as
\begin{equation}
\nabla_\mu (\rho_{\rm i} u^\mu) =S_{\rm i} \ \ ({\rm i = p, \alpha, C}),
\label{s-8}
\end{equation}
where $S_{\rm i}$ is the source term due to the nuclear burning for each nucleus. 
The baryon number density should be conserved, and hence, $S_{\rm i}$ should fulfill the equation
\begin{equation}
S_{\rm p}  + S_{\rm \alpha} + S_{\rm C} = 0.
\end{equation}
Our method for numerically solving Eq.~(\ref{s-8}) will be described in Sec.~\ref{nuc}.
\subsection{Equation of State}
We assume that the equation of state~(EOS)  during the collapse can be written as a sum of the ideal gas and radiation, i.e.,
\begin{equation}
P = \frac{1}{3}aT^4 + \frac{Y_{\rm T}k_{\rm B} }{m_{\rm u}} \rho T,
\label{s-9}
\end{equation}
\begin{equation}
\epsilon = \frac{aT^4}{\rho} + \frac{3}{2} \frac{Y_{\rm T}k_{\rm B} }{m_{\rm u}}  T,
\label{s-10}
\end{equation}
where $a, k_{\rm B}$, and $T$ are the radiation constant, Boltzmann constant, and temperature, respectively. 
$Y_{\rm T}$ is defined by
\begin{equation}
Y_{\rm T} = Y_{\rm p} + Y_{\rm \alpha} + Y_{\rm C} + Y_{\rm e}.
\label{s-11}
\end{equation}
We write the mass fraction of each nucleus as $X_\alpha = 4Y_\alpha$ and $Z_{\rm C}= 12Y_{\rm C}$. 
If a SMS core is in the ZAMS phase, their typical values are approximately $(Y_{\rm p},~X_{\alpha},~Z_{\rm C}) \approx (0.75,~0.25,~0)$,
leading to $Y_{\rm T}\approx 1.69$ and if a SMS core is just at the onset of the helium burning phase, $(Y_{\rm p},~X_{\alpha},~Z_{\rm C}) \approx (0,~1,~0)$, and thus, $Y_{\rm T} \approx 0.75$.

The equations of state are valid unless electrons become relativistic, degenerate, or the effects of pair creation of electron-positron pairs cannot be neglected. 
That is, this approximation is valid within the range of $\rho < 10^6~{\rm g/cm^3}$ and $T < 10^9~{\rm K}$. 
Actually, just before the central black hole is formed, the central region of the collapsing
SMS core would be out of this range but our simulations show that this region immediately falls into the black hole.
Hence it will be safe to consider that the EOS composed of Eqs.~(\ref{s-9}) and (\ref{s-10}) is appropriate for our present study. 

\subsection{Nuclear burning}
\label{nuc}
By using $\nabla_\mu (\rho u^\mu) =0$, Eq.~(\ref{s-8}) can be rewritten as
\begin{equation}
\frac{\partial{Y_{\rm i}}}{\partial{t}}+v^j\frac{\partial{Y_{\rm i}}}{\partial{x^j}}=\frac{S_{\rm i}}{\rho u^t},
\label{eqy}
\end{equation} 
where $v^j = u^j/u^t$. 
It is not an easy task to simultaneously solve the advection and the nuclear burning network. 
Thus we divide Eq.~(\ref{eqy}) into two parts~(i.e., operator splitting approach is employed). 
First, we solve the advection equations without nuclear burning, that is, 
\begin{equation}
\frac{\partial{Y_{\rm i}}}{\partial{t}}+v^j\frac{\partial{Y_{\rm i}}}{\partial{x^j}}=0.
\label{eqyad}
\end{equation} 
This is equivalent to solving
\begin{equation}
\nabla_\mu (\rho_{\rm i} u^\mu) =0 \ \ ({\rm i = p, \alpha, C}).
\label{eqr}
\end{equation}
Second, we solve equations of nuclear burning reactions such as
\begin{eqnarray}
\frac{\partial{Y_{\rm p}}}{\partial{t}} &=& \frac{S_{\rm p}}{\rho u^t}=\frac{m_{\rm u}}{u^t}  \left (- \frac{ \dot{q}_{\rm CNO}}{Q_{\rm CNO}} \right ),\\
\frac{\partial{Y_{\rm \alpha}}}{\partial{t}} &=&\frac{S_{\rm \alpha}}{\rho u^t}= \frac{m_{\rm u}}{4u^t}\left (  \frac{\dot{q}_{\rm CNO}}{Q_{\rm CNO}} -\frac{\dot{q}_{3\alpha}}{Q_{3\alpha}} \right ), \\
\frac{\partial{Y_{\rm C}}}{\partial{t}} &=&\frac{S_{\rm C}}{\rho u^t}= \frac{m_{\rm u}}{12u^t}\left ( \frac{\dot{q}_{3\alpha}}{Q_{3\alpha}} \right ), 
\end{eqnarray}
where $\dot{q}_{\rm J}$ and ${Q}_{\rm J}~({\rm J=CNO, 3\alpha})$ are the energy generation rate and the energy liberated per baryon of CNO cycle and triple-alpha reaction, respectively. $\dot{q}_{\rm J}$ has units of erg/g/s.
In this paper, we only consider cold CNO cycle, hot CNO cycle, and triple-alpha reactions. 
For $\dot{q}_{\rm J}$, we employ the same formulae as those of Ref.~\cite{2012ApJ...749...37M} (see Eqs.~(25--27) for this reference). 
We also simulated the gravitational collapse including the effect of the rp-process as test calculations and found that 
the rp-process affects only weakly the gravitational collapse because this process becomes efficient only in the very dense and hot region. 
Thus, for the product runs, we neglect the rp-process. 
We briefly illustrate the validity of our calculation of nuclear burning in Appendix~\ref{app1}.

We also compute the neutrino generation rate using the formulation of~Ref.~\cite{1996ApJS..102..411I} (see Eqs.~(2.1),~(3.2), and (4.1) for this reference), but we do not include the effect of neutrino cooling in our simulations because it is much weaker than the effect of nuclear burning outside the formed black hole except just before the black-hole formation (see Sec.~\ref{nuclear}). At the black-hole formation, the neutrino cooling rate would become larger than the nuclear burning heating rate. However at this time, most of the generated neutrinos would be absorbed  by the black hole. Thus, we assume that the neutrino cooling would be negligible throughout the collapse.

\subsection{Initial conditions}
Following our previous paper~\cite{0004-637X-818-2-157}, we first prepare the equilibrium state of SMS cores which are marginally stable to the general-relativistic quasi-radial instability. We briefly review the method as follows. 

{
In the stationary and axisymmetric spacetime, the metric can be written by
\begin{eqnarray}
ds^2 = -e^{\gamma_{\rm s}+\rho_{\rm s}} c^2 dt^2  +e^{2\alpha_{\rm s}} (dr^2 + r^2 d\theta^2) \\ \nonumber
 + e^{\gamma_{\rm s}-\rho_{\rm s}} r^2 {\rm sin}^2 \theta (d\varphi -\omega_{\rm s}  dt)^2, \label{metc}
\end{eqnarray}
where $\rho_{\rm s},~\gamma_{\rm s},~\alpha_{\rm s}$, and $\omega_{\rm s}$ are functions of $r$ and $\theta$. 
}

The EOS employed is the same as Eqs.~(\ref{s-9}) and (\ref{s-10}). Using the first law of thermodynamics,  
the adiabatic constant $\Gamma$ is calculated as~\cite{1984ApJ...280..825B}
\begin{equation}
\Gamma = \left ( \frac{\partial   {\rm ln}P}{\partial {\rm ln} n_{\rm B} }\right )_s = \frac{4}{3} + \frac{4\sigma+1}{3(\sigma+1)(8\sigma+1)},
\label{s-12}
\end{equation}
where $\sigma$ is the ratio of the radiation pressure to the gas pressure defined by
\begin{equation}
\sigma \equiv \frac{a T^3}{3Y_{\rm T}n_{\rm B} k_{\rm B}} = \frac{s_\gamma}{4Y_{\rm T} k_{\rm B}}.
\label{s-13}
\end{equation}
Here $s_\gamma$ and $s = s_\gamma + s_{\rm g}$ are the photon entropy per baryon and the total (photon and gas) entropy per baryon, respectively. 
If SMS cores are in nuclear-burning phases, they should be fully convective (e.g.,~Refs.~\cite{1984ApJ...280..825B, 2016ApJ...830L..34U}). 
Hence, it is natural to assume that SMS cores are isentropic~($s=$ constant), its chemical composition is uniform~($Y_I =$ constant), and they are rigidly rotating. Furthermore in SMS cores, $s_\gamma \gg s_{\rm g}$ are realized~\cite{1984ApJ...280..825B, 1992ApJ...398..203C} so that $s_\gamma$ is also nearly constant. 
Then $\sigma$ can be assumed to be constant. As a result, Eq.~(\ref{s-12}) can be easily integrated by $n_{\rm B}$, and we get the polytroics EOS
\begin{equation}
P=K\rho_0^{\Gamma} , \ \ \Gamma=1+\frac{1}{N},
\label{s-14}
\end{equation}
where $K$ and $N$ are the polytropic constant and the polytropic index, respectively. $K$ is written as
\begin{equation}
K \approx \left ( \frac{Y_{\rm T}k_{\rm B} \sigma}{m_{\rm u}} \right )^\frac{4}{3} \left ( \frac{ 3 }{a  } \right )^\frac{1}{3} (1+\sigma^{-1}) \rho_0^{-{1}/{(6\sigma)}}.
\label{s-15}
\end{equation}
Here, the density dependence of $K$ can be neglected because $\sigma \gg 1$ for typical SMS cores. 
{We use this polytropic EOS in computing the equilibrium state and assume that the energy momentum tensor has the same form as Eq.~(\ref{s-4}).}

For SMS cores, their density profile is approximated by the Lane-Emden solution of $N=3$ even if 
they are rotating at mass-shedding limit~(e.g.,~Ref.~\cite{1971reas.book.....Z}). 
Then by using this solution, the mass of the SMS core, $M$, can be approximately written as
\begin{equation}
M \approx 4.555 G^{-\frac{2}{3}} K^\frac{3}{2}.
\label{s-15.1}
\end{equation}
(This is equivalent to approximating $C_{n_p}=C_3$ of Ref.~\cite{0004-637X-818-2-157}.)

Using Eqs.~(\ref{s-12}), (\ref{s-14}), (\ref{s-15}), and (\ref{s-15.1}), $\Gamma$ can be approximately rewritten as
\begin{equation}
\Gamma- \frac{4}{3} \approx \frac{1}{6\sigma} \approx 3.8\times 10^{-3} \left (   \frac{  M }{ 10^5M_\odot  }  \right )^{-\frac{1}{2}} \left (   \frac{  Y_{\rm T} }{1.69   }  \right ),
\label{s-15.2}
\end{equation}
where we used $\sigma \gg 1$.
Inserting Eq.~(\ref{s-13}) to Eq.~(\ref{s-15.2}), the relation among the central density, the central temperature, and the mass of the SMS core can be approximately written as  
\begin{equation}
\frac{P_{\rm c}}{\rho_{\rm 0c}c^2} \sim 1. 1 \times 10^{-3} \left ( \frac{ T_{\rm c} }{10^{8. 2}K  } \right )\left ( \frac{  M}{  10^5 M_\odot } \right )^\frac{1}{2},
\label{s-16}
\end{equation}
where the index ${\rm c}$ denotes the central value of the SMS core. 

We define two dimensionless parameters, $y$ and $\beta$, by
\begin{equation}
y \equiv 2. 6324\frac{P_{\rm c}}{\rho_{\rm 0c}c^2},
\label{s-17}
\end{equation}
and
\begin{equation}
\beta \equiv \frac{T_{\rm rot}}{|W|}.
\label{s-18}
\end{equation}
Here、$T_{\rm rot}$ and $W$ denote the rotational kinetic energy and the gravitational potential energy {defined by
\begin{eqnarray}
T_{\rm rot} &\equiv& \frac{1}{2}J_{\rm K}\Omega \label{ktots},\\
W &\equiv& M_{\rm K}c^2 -M_{\rm p}c^2 - T_{\rm rot},
\end{eqnarray}
where $\Omega$, $J_{\rm K}$, $M_{\rm K}$, and $M_{\rm p}$ are the angular velocity, Komar angular momentum, Komar mass (gravitational mass), and the proper mass defined by
\begin{eqnarray}
J_{\rm K} &\equiv& 2\pi \int \rho_0 h u^t u_\varphi {\rm e}^{2\alpha_{\rm s} + \gamma_{\rm s}} r^2 dr d{\rm cos}\theta,\\
M_{\rm K} &\equiv& \frac{2\pi}{c^2} \int (-2T^t_t + T^\mu_\mu) {\rm e}^{2\alpha_{\rm s} + \gamma_{\rm s}} r^2 dr d{\rm cos}\theta,
\end{eqnarray}
and 
\begin{equation}
M_{\rm p} \equiv \frac{2\pi}{c} \int \rho_0 u^t (c^2+\epsilon) {\rm e}^{2\alpha_{\rm s} + \gamma_{\rm s}} r^2 dr d{\rm cos}\theta,
\end{equation}
} 
respectively. 

According to our previous result~\cite{0004-637X-818-2-157}, the condition for SMS cores to be marginally stable to gravitational collapse 
can be written as
\begin{equation}
\Gamma - \frac{4}{3} = y - y^2 - \left ( \frac{10}{3} - 2\Gamma -y -\beta \right ) \beta.
\label{s-19}
\end{equation}
If the value of the left-hand side of Eq.~(\ref{s-19}) is smaller than the right-hand side, the SMS core is unstable. 
Using Eqs.~(\ref{s-15.2})--(\ref{s-19}), the mass of the SMS core which is marginally stable can be approximately written as
\begin{equation}
M_5^\frac{1}{2} = \frac{\beta_{\rm -3} +\sqrt{\beta_{\rm -3}^2 + 99 T_{8. 2}Y_{T 1. 69}}}{8. 7 T_{8. 2}},
\label{s-20}
\end{equation}
where 
\begin{eqnarray}
M_5 \equiv \left (   \frac{  M }{10^5M_\odot   }  \right ), \ T_{8. 2} \equiv \left (   \frac{   T_{\rm c}}{ 10^{8. 2} {\rm K} }  \right ), \nonumber \\ 
\ Y_{T 1. 69} \equiv \left (   \frac{ Y_{\rm T}  }{1. 69   }  \right ), \ \beta_{\rm -3} \equiv \left (   \frac{   \beta}{ 10^{-3}  } \right ).
\end{eqnarray}

If a SMS core is in nuclear burning phases, its surface luminosity, $L$, and its energy generation rate of nuclear burning, $\dot{Q}$ should agree with each other. Here, $L$ and $\dot{Q}$ are defined by
\begin{eqnarray}
L &\equiv&  - \frac{4ac}{3\kappa} \int_S \frac{1}{\rho_0} {\nabla}^i (T^4)  d{S_i}, \label{s-20.1} \\
\dot{Q} &\equiv& \int_{V_S} \rho_0 \dot{q} dV \label{s-20.2} ,
\end{eqnarray}
where $\kappa,~S$, and $V_S$ denote the opacity, the surface and the volume of the SMS core, respectively. We assume that the opacity is dominated by Thomson scattering of free electrons, i.e., $\kappa = 0.4 Y_{\rm e}$~${\rm cm^2/g}$.

We use an iteration method to derive the equilibrium state of marginally stable SMS cores using Eqs.~(\ref{s-13}), (\ref{s-19}), (\ref{s-20.1}), and (\ref{s-20.2}) as follows. 
The input parameters are the chemical composition $Y_{\rm I}~({\rm I=p, \alpha, C})$ and the rotation parameter $\beta$. 
\begin{enumerate}
	\item Provide input parameters $Y_{\rm I}~({\rm I=p, \alpha, C})$ and $\beta$. 
	\item Provide a temporal value of $y$. 
	\item Calculate the adiabatic constant $\Gamma$ by using Eq.~(\ref{s-19}). Then by using Eqs.~(\ref{s-12}), (\ref{s-13}) and (\ref{s-15}), we determine the polytropic constant $K$. 
	\item Calculate { $\rho_{\rm s},~\gamma_{\rm s},~\alpha_{\rm s},~\omega_{\rm s}$}, and the density profile of the SMS core by solving the set of equations for stationary axisymmetric rotating equilibrium in general relativity by using the method of Ref.~\cite{1992ApJ...398..203C, 1992ApJ...398..203C}.
	\item Calculate the temperature profile from Eq.~(\ref{s-13}). 
	\item Calculate $\dot{Q}$ and $L$. If $\dot{Q}>L$, we decrease $y$, and otherwise, we increase $y$, and return to step 3. 
\end{enumerate}
We iteratively perform the procedures 1--6 untill $|\dot{Q}-L|$ becomes sufficiently small. 
	
In numerical simulations, we initially reduce the temperature uniformly as the initial perturbation.
We define a perturbation parameter  $D_{\rm T} $ such that the perturbed temperature can be written as 
\begin{equation}
T=\left (1-\frac{D_{\rm T}}{100}\right )T_0,
\label{s-20.3}
\end{equation}
where $T_0$ is the unperturbed temperature. We peform simulations for $D_{\rm T}=0.5,~0.25,~0.125$ and $0.0625$, respectively
~(this is approximately the same as uniformly reducing the pressure by $2\%,~1\%,~0.5\%$, and $0.25\%$, respectively). 

After adding the perturbation, the configuration does not satisfy the constraint equations, and hence, we once more solve the constraint equations. 

\section{Result}
\label{result}
\subsection{Overview of the collapse}
\label{over}
\begin{table}[htpb]
	\caption{Key quantities for SMS cores employed in this paper. $M_5 \equiv M_0/10^5M_\odot$. "Shedd" means that the SMS core is at mass-shedding limit~($ T_{\rm rot}/|W|\approx 0.009$). $Z_{\rm C9}$ is defined by $Z_{\rm C}/10^{-9}$. $\Gamma$ is the initial polytropic index defined by Eq.~(\ref{s-14}).}
	\begin{tabular}{|c||c|c|c|c|c|c|c|} \hline
		Model & $M_5$ & $T_{\rm rot}/|W|  $ & $Y_{\rm p}$ & $X_{\rm \alpha}$ & $Z_{\rm C9} $ & ${\Gamma}$ & phase \\ \hline \hline
		A1 & 1.99 & 0.002 & 0.75& 0.25& $5$ & 1.3360 & ZAMS\\ \
		A2 & 0.47 & 0.002 & 0& 1.0& 0 & 1.3358 & He-burning\\ \
		A3 & 6.56 & Shedd & 0.75& 0.25& $5$ &1.3348 & ZAMS\\ \
		A4 & 1.57 & Shedd & 0&1.0& 0 &1.3347 & He-burning\\  \hline
	\end{tabular}
	\label{t-1}
\end{table}

\begin{figure}[htpb]
\includegraphics[width=1\linewidth]{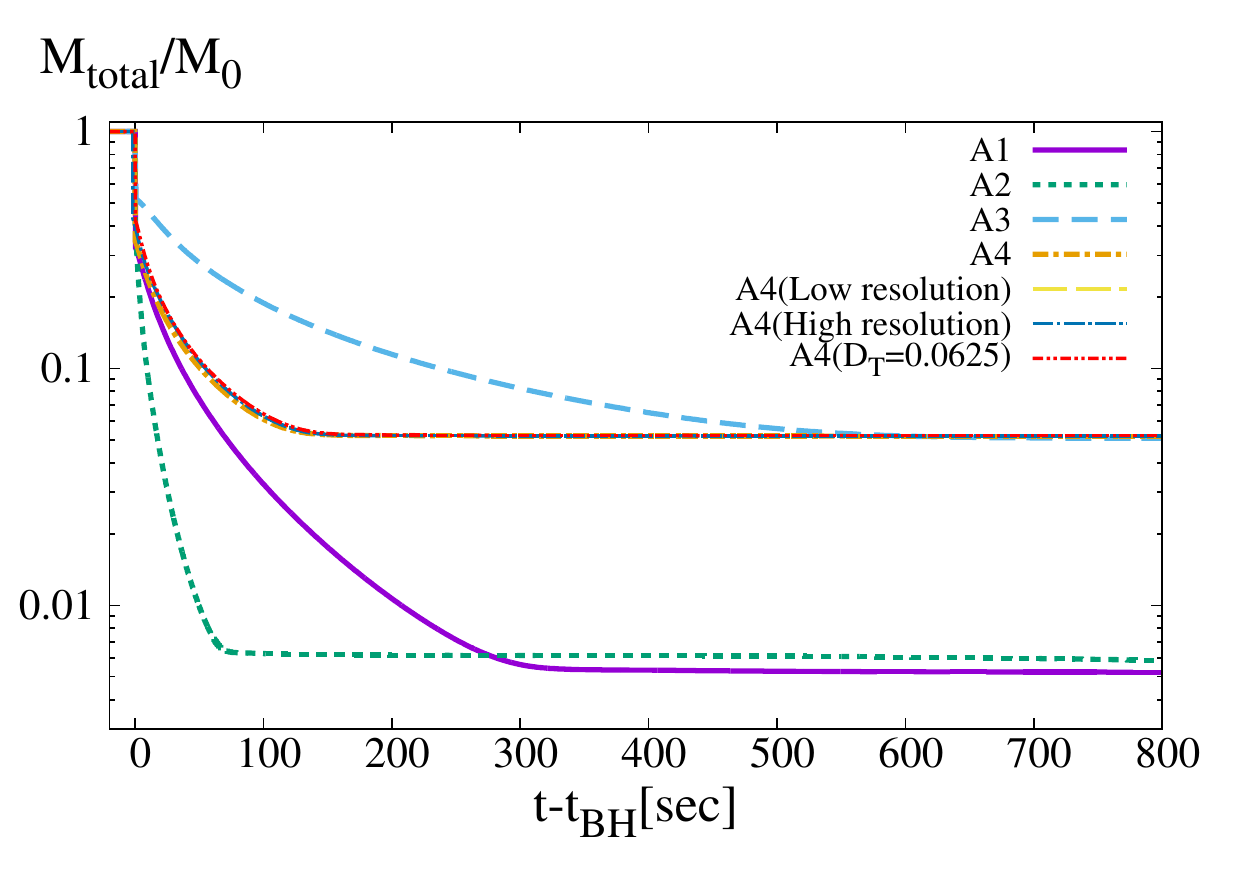}
	\caption{Time evolution of the total mass located outside the formed black hole. The purple-solid, green-dotted, light-blue-dashed, and orange-dashed-dotted curves denote the results for models A1--A4 with $D_{\rm T}=0.5$, respectively. The yellow-long-dashed and blue-long-dashed-dotted curves are for model A4 with low and high grid resolutions, respectively. The red-dashed-dotted-dotted curve is for model A4 with $D_{\rm T}=0.0625$. $t_{\rm BH}$ is the time of the black-hole formation.}
		\label{fig:mass14}
	\end{figure}

\begin{figure}[htpb]
	\includegraphics[width=1\linewidth]{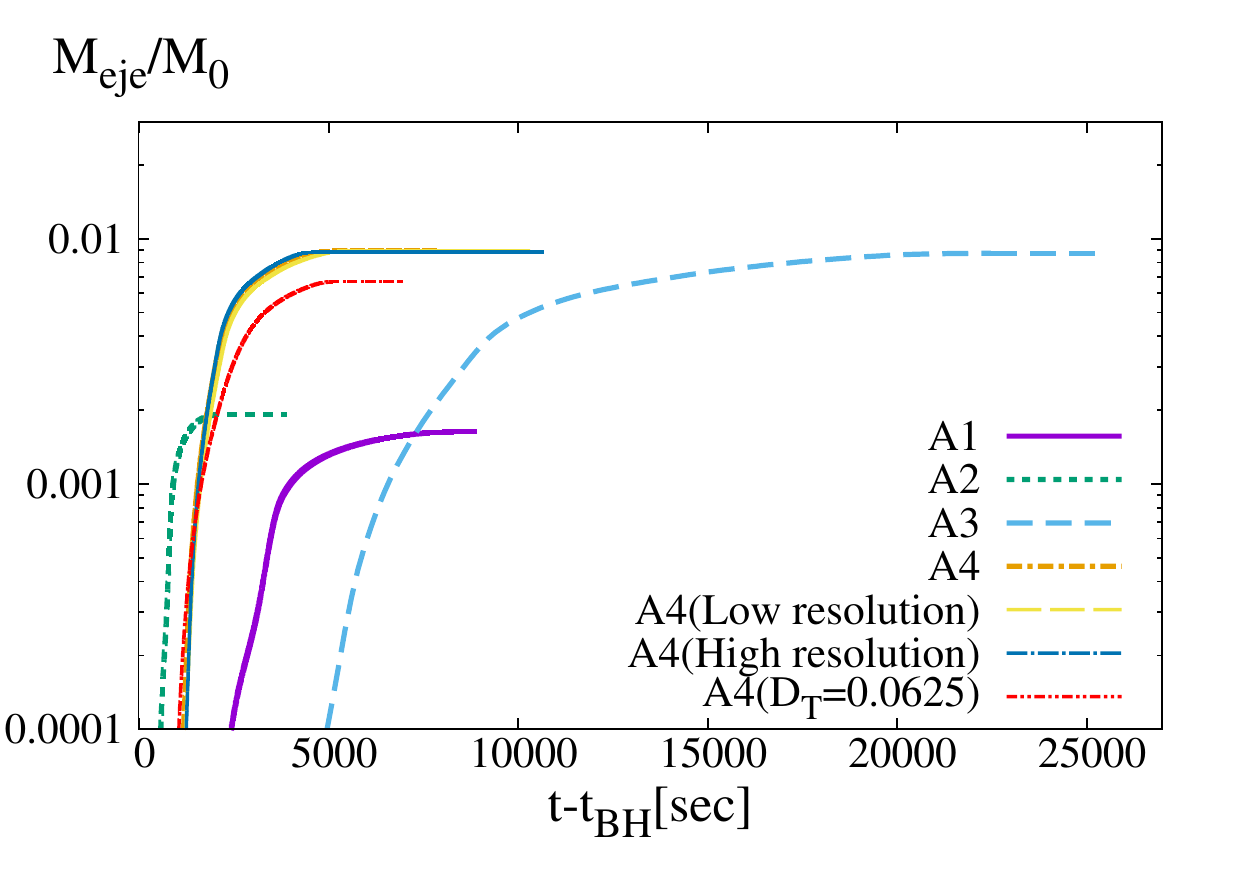}
	\caption{Time evolution of $M_{\rm eje}(D,t)$, the unbound mass ejected from the domain of $\sqrt{X^2+Z^2}<D$ defined by Eq.~(\ref{ejected}). We take $D=600R_{\rm M}$. The purple-solid, green-dotted, light-blue-dashed, and orange-dashed-dotted curves denote the results for models A1--A4 with $D_{\rm T}=0.5$, respectively. The yellow-long-dashed and blue-long-dashed-dotted curves are for model A4 with low and high grid resolutions, respectively. The red-dashed-dotted-dotted curve is for model A4 with $D_{\rm T}=0.0625$. $t_{\rm BH}$ is the time of the  black-hole formation.}
	\label{fig:flout14}
\end{figure}

We performed numerical relativity simulations for four initial states of SMS cores listed in Table \ref{t-1}.
Basically, the simulations are performed for the middle grid resolution with $D_{\rm T}=0.5$. 
For selected models, the simulations are performed for different grid resolutions and for different values of $D_{\rm T}$ (see, e.g., Figs.~\ref{fig:mass14} and \ref{fig:flout14}).
Models A1 and A3 are assumed to be in the ZAMS phase, and A2 and A4 are just at the onset of the helium-burning phase.  

\begin{figure*}[htpb]
	\begin{center}
	\includegraphics[trim=0 0 0 0, width=1\linewidth]{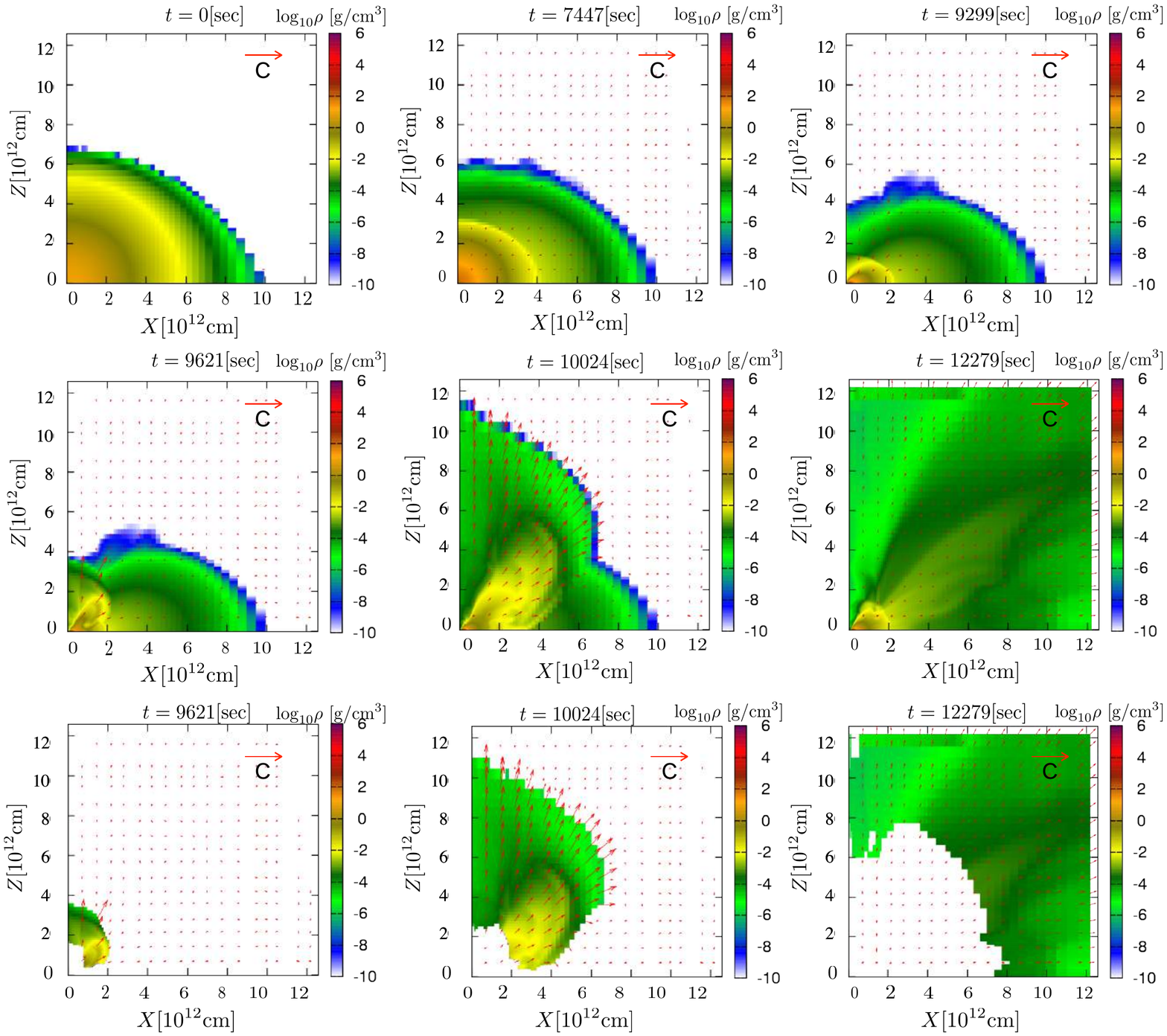}
	\end{center}
	\caption{Snapshots of density profiles for the SMS core collapse for model A4. The 7th--9th panels show only unbound material. The red arrows denote the velocity profile, $u^i/u^t~(i=X,Z)$, which are normalized as indicated in the upper right-hand corner of each snapshot.}
	\label{fig:anime}
\end{figure*}

Figure~\ref{fig:anime} displays snapshots of the rest-mass density profiles for the collapse of a SMS core to a black hole and a torus surrounding it for model A4. 
The lower panels are the snapshots at the same time as the middle panels but they only depict unbound fluid elements. 
The black hole is formed at $t \approx 9400$ s~(near the time of the 3rd panel). After the collapse, a torus surrounding the black hole is formed and a part of the mass becomes unbound and ejected. Qualitatively, the collapse dynamics for the other models is similar to model A4.

Figure~\ref{fig:mass14} shows the evolution of the total mass outside the central black hole, $M_{\rm total}(t)$, defined by 
\begin{equation}
M_{\rm total}(t) \equiv \int_{V'} \rho_{0*} dV,
\label{mtotal}
\end{equation}
as a function of $t-t_{\rm BH}$ where $t_{\rm BH}$ is the time of the black-hole formation. 
In Eq.~(\ref{mtotal}), $V'\equiv V- V_{\rm BH}$. $V$ and $V_{\rm BH}$ are the 3D spatial volume of $t={\rm const}$ and the region inside the black hole, respectively. Before the black-hole formation, $V'=V$.
$\rho_{0*} \equiv \rho_0 \sqrt{-g}c u^t$ and $dV \equiv 2\pi dX dZ$ are the weighted rest-mass density and the volume element, respectively.
We note that $t_{\rm BH}$ for models A1,~A2,~A3 and A4 with $D_{\rm T}=0.5$ are 28664~s, 7311~s, 38243~s, and 9402~s, respectively. 
The black-hole formation time, $t_{\rm BH}$, for model A4 with $D_{\rm T}=0.0625$ is 109895 s, and hence, $t_{\rm BH}$ depends strongly on the initial perturbation. (However, the final outcomes of the collapse depend only weakly on the initial perturbation: see below.)  

After the black-hole formation, the mass accretes to the black hole and the accretion terminates at $t-t_{\rm BH}\approx 300~{\rm s},~70~{\rm s},~800~{\rm s}$, and $150$ s for models A1--A4, respectively. 
To specify the mass which is not absorbed by the black hole,
we define $M_{\rm ter}$ as the value of $M_{\rm total}$ at the time at which the accretion to the black hole terminates. 
We list $M_{\rm ter}$ together with the mass ($M_{\rm BH}$) and dimensionless spin parameter ($a_{\rm BH}$) of the final state of the black holes in Table~\ref{t-3}. 
Approximately $0.5\%$ of the initial mass is located outside the black hole at the final state for models A1 and A2. On the other hand, these values are $5\%$ for rapidlly rotating  models A3 and A4, respectively. The values of the dimensionless spin parameter, $a_{\rm BH}$ are $\approx 0.5$ for models A1 and A2 and $\approx 0.7$ for models A3 and A4, respectively.

We find that the accretion for hydrogen-burning models A1 and A3 occurs more slowly than helium-burning models A2 and A4. 
This is caused by the fact that the SMS core density (and spacetime curvature) for models A1 and A3 are lower than for models A2 and A4. 
The curves for model A4 with $D_{\rm T}=0.5$ and $0.0625$ show that 
the property of the accretion of the mass to the formed black hole depends weakly on $D_{\rm T}$. 
This is due to the fact that after the formation of the black hole, the accreting matter is approximately in free fall, and thus, the property of the accretion of the mass does not depend strongly on the initial perturbation.

Figure~\ref{fig:flout14} shows the time evolution of the mass ejected from the domain $\sqrt{X^2+Z^2} < D$, $M_{\rm eje}(D,t)$, defined by Eq.~(\ref{ejected}). We take $D=600R_{\rm M}$ and define $M_{\rm eje}(D) \equiv M_{\rm eje}(D,t=t^*)$ where $t^*$ is the time at which all of the fluid elements of the outflow finishes escaping from the domain $\sqrt{X^2+Z^2}< D$. 
We find that $M_{\rm eje}(D)$ depends only weakly on $D$, and thus, we omit the argument $D$ in $M_{\rm eje}$ in the following.
It is found that 
$M_{\rm eje}/M_0$ is approximately $0.2\%$ for models A1 and A2 and $1\%$ for rapidly rotating models A3 and A4, respectively.  
Thus $M_{\rm eje}$ is approximately $1/5$ times smaller than $M_{\rm ter}$. 

Unlike $M_{\rm ter}$, the total mass of the outflow for model A4 with $D_{\rm T}=0.0625$ is approximately $0.7$ times as large as  model with $D_{\rm T}=0.5$. We will discuss the reason for this as well as the property of the outflow in detail in Sec.~\ref{outflow} (see also Appendix \ref{initp}. for a method to take $D_{\rm T}=0$ limit for the outflow). 

Before closing this subsection, we note the convergence property of the numerical results.
Figures~\ref{fig:mass14} and \ref{fig:flout14} show a good convergence of the numerical result for model A4 among the low, middle, and high resolution results. 
We check that for models A1 and A4, the values of $M_{\rm total}$ at $t-t_{\rm BH}=2000$ s and $M_{\rm eje}$ agree with each other among the low, middle, and high-resolution cases within $1.5\%$ disagreement.


	\begin{table*}[htpb]
	\caption{Quantities for the gravitational collapse of SMS cores. $M_{05}$: initial mass of SMS cores in units of $10^5M_\odot$. $M_{\rm BH5}$ and $a_{\rm BH}$: mass in units of $10^5M_\odot$ and spin of the remnant black hole. $M_{\rm ter3}$: The total mass located outside the black hole after the termination of the mass accretion to the black hole in units of $10^3M_\odot$. $E_{\rm tor}$: internal energy of the torus. $\dot{E}_{\rm nuc}$ and $\dot{E}_{\nu}$: nuclear energy generation rate and neutrino generation rate of the torus. $T_{\rm max}$ and $\rho_{\rm max}$ : the maximum temperature and density of the torus. $\tau_{\rm nuc}$ and $\tau_{\rm dyn}$: heating timescales by nuclear burning and dynamical time (rotation period at the density maximum of the torus). These values are calculated at $t-t_{\rm BH} = 2000$ s.}
	\begin{tabular}{|c||c|c|c|c|c|c|c|c|c|c|c|c|} \hline
		Model & $M_{05}$ & $M_{\rm BH5}$ & $a_{\rm BH}/M_{\rm BH}$ &$ M_{\rm ter3}$&$E_{\rm tor}$[erg] & $ \dot{E}_{\rm nuc}$[erg/s] & $\dot{E}_{\nu}$ [erg/s] & $T_{\rm max}[10^9{\rm K}]$ & $\rho_{\rm max}[{\rm g/cm^3}]$ & $\tau_{\rm nuc}$[s] & $\tau_{\rm dyn}$[s] \\ \hline \hline
		A1 & 1.99 & 1.98   & 0.50 &1.0 &$1\times 10^{55}  $& $1\times 10^{45}$ & $1\times 10^{37}$ & 4.2 & 20   &$1 \times 10^{10} $&110   \\ \
		A2 & 0.470 & 0.467 & 0.51 &0.29 &$4\times 10^{54}$& $4\times 10^{46}$ & $3\times 10^{43}$ & 9.3 & 400 &$9\times 10^{7}   $&25   \\ \
		A3 & 6.56 & 6.23   & 0.69 & 33&$7\times 10^{56}$& $1\times 10^{46}$ & $1\times 10^{43}$ & 5.2 & 22   &$7\times 10^{10} $& 190 \\ \
		A4 & 1.57 & 1.49   & 0.69 & 8.2&$2\times 10^{56}$& $1\times 10^{48}$ & $3\times 10^{47}$ & 10  & 400 &$2\times 10^{8}  $ &46   \\ \hline
	\end{tabular}
	\label{t-3}
\end{table*}
	\subsection{Nuclear and neutrino interaction rate}
	\label{nuclear}
	\begin{figure}[htpb]
	\includegraphics[width=1\linewidth]{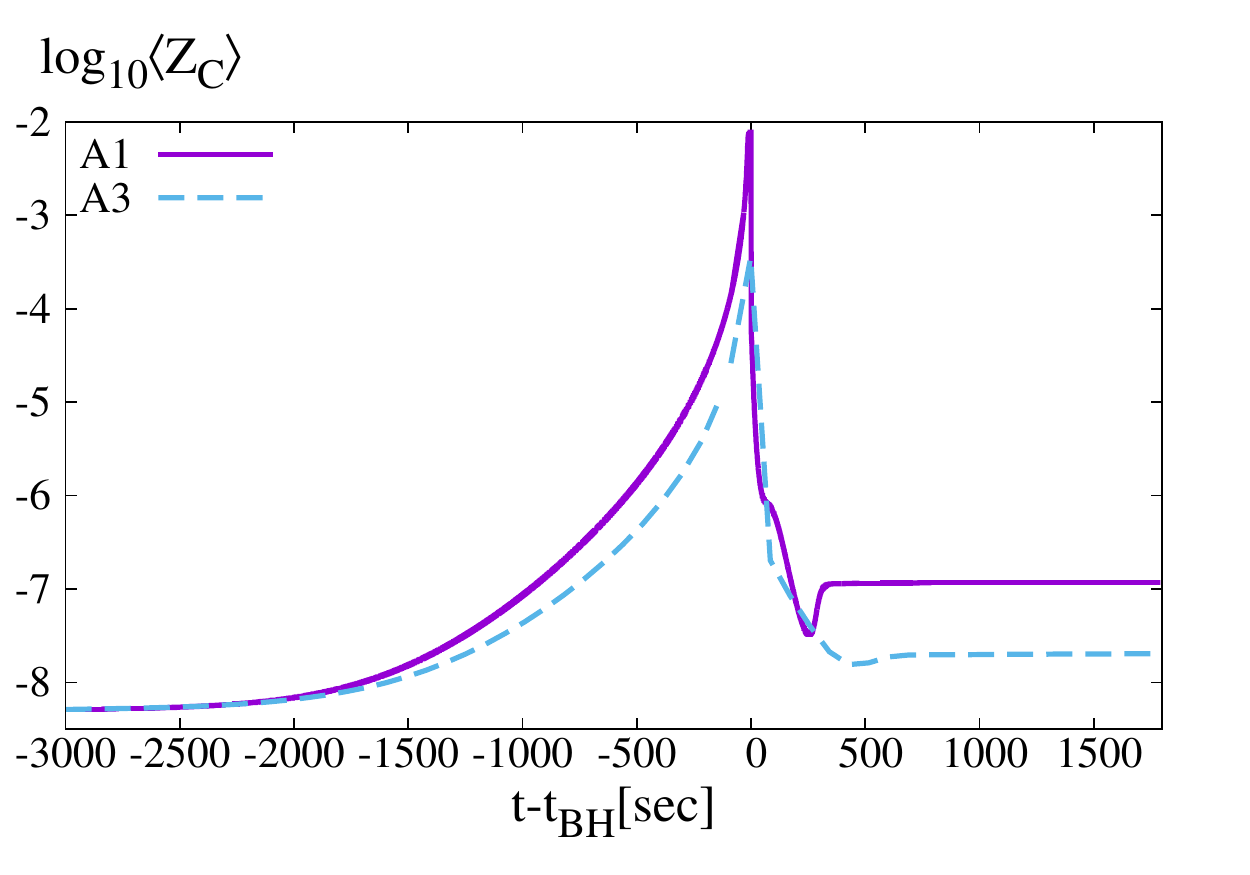}
	\caption{Time evolution of the averaged mass fraction of carbon for the matter located outside the black hole for hydrogen-burning models A1 (purple-solid) and A3 (light-blue-dashed). }
	\label{fig:A13}
\end{figure}
\begin{figure}[htpb]
	\includegraphics[width=1\linewidth]{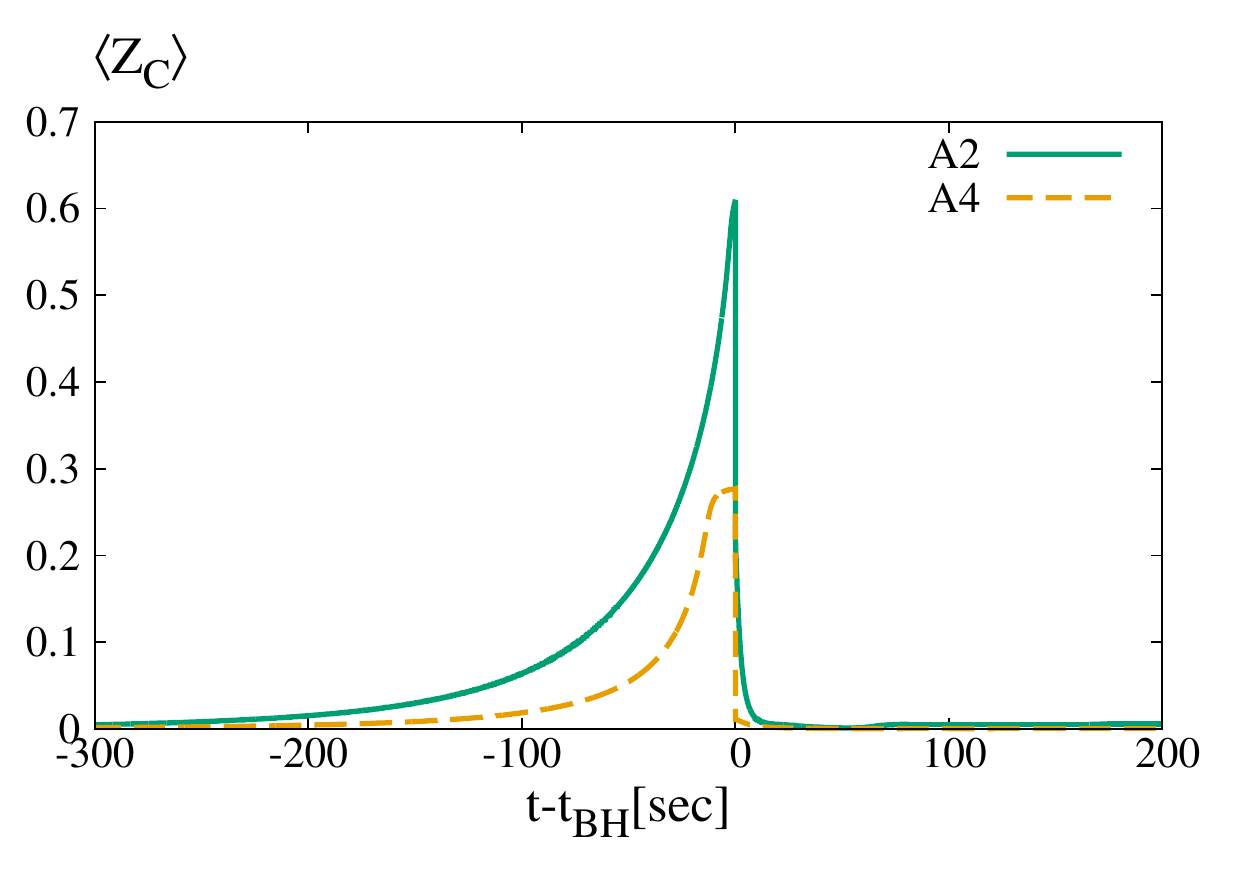}
	\caption{Time evolution of the averaged mass fraction of carbon for the matter located outside the black hole for helium-burning models A2 (green-solid) and A4 (orange-dashed). }
	\label{fig:A24}
\end{figure}
\begin{figure}[htpb]
	\includegraphics[angle=0, width=1\linewidth]{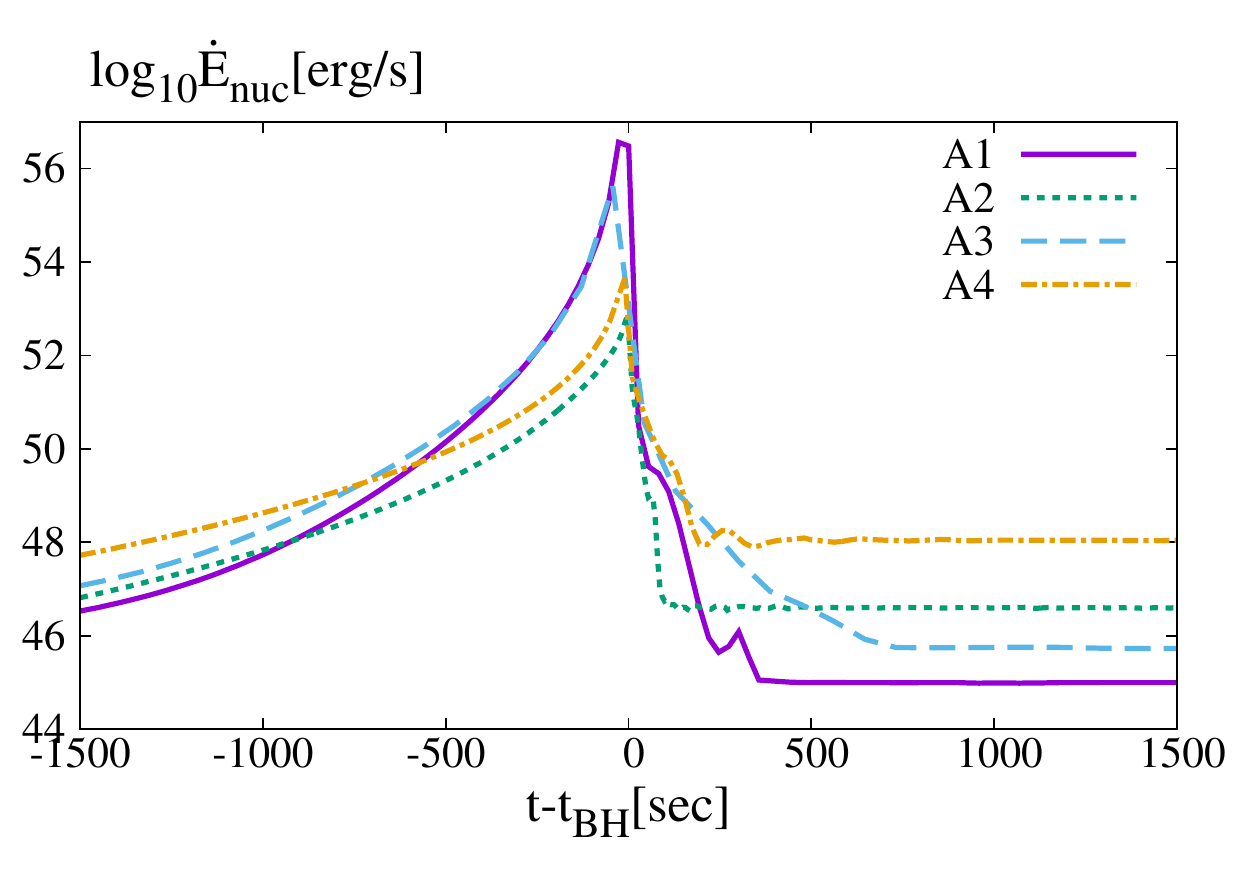}
	\caption{Time evolution of the total energy generation rate of nuclear burning for the matter located outside the black hole for models A1 (purple-solid), A2 (green-dotted), A3 (light-blue-dashed), and A4 (orange-dashed-dotted), respectively.}
	\label{fig:nuc14}
\end{figure}
\begin{figure}[htbp]
	\includegraphics[angle=0, width=1\linewidth]{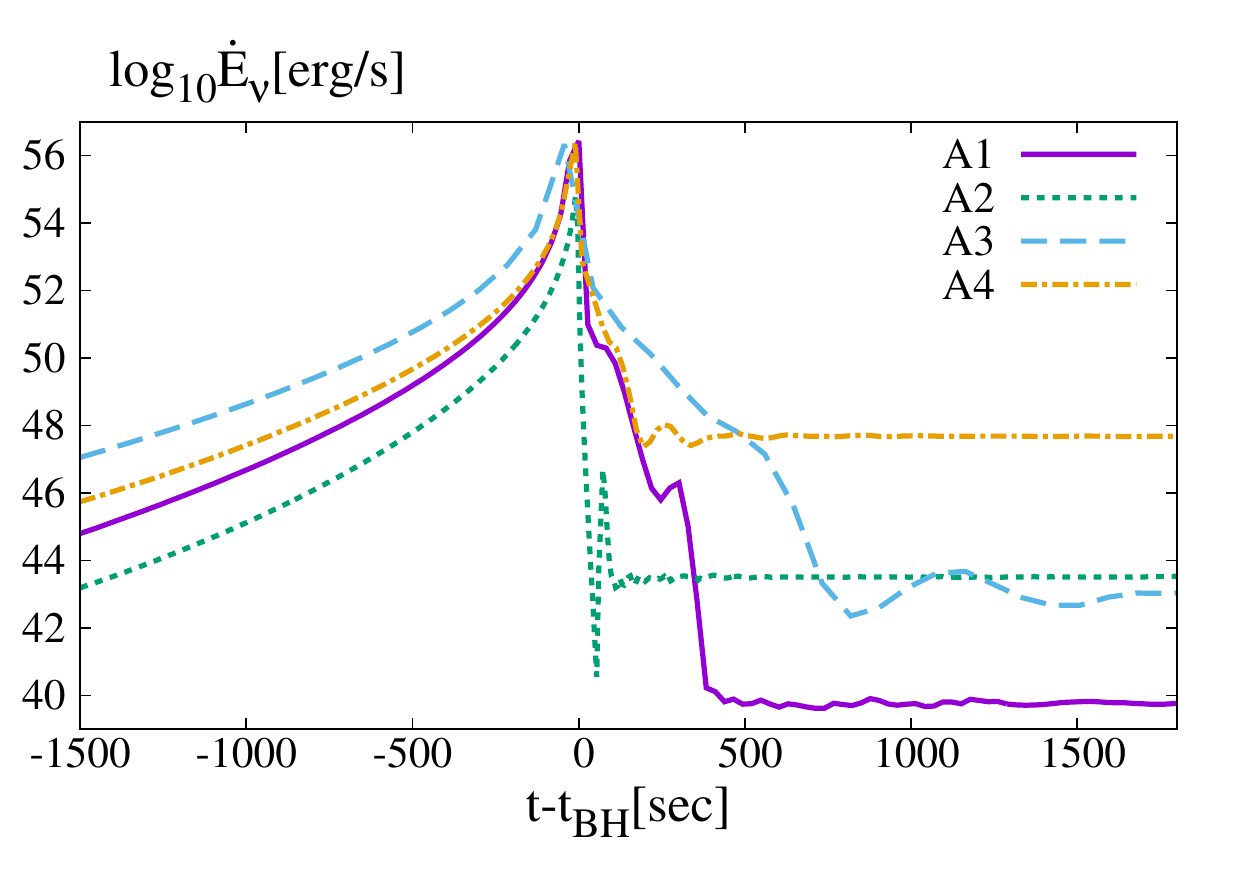}
	\caption{Time evolution of the neutrino generation rate for the matter located outside the black hole for models A1 (purple-solid), A2 (green-dotted), A3 (light-blue-dashed), and A4 (orange-dashed-dotted), respectively.}
	\label{fig:neu14}
\end{figure}
Figures~\ref{fig:A13} and \ref{fig:A24} display the time evolution of the averaged mass fraction of $^{12}$C defined by $\langle Z_{\rm C}\rangle = M_{\rm C}/M_{\rm total}$ for hydrogen-burning models A1 and A3~(Fig.~\ref{fig:A13}) and helium-burning models A2 and A4~(Fig.~\ref{fig:A24}), respectively.
Here, $M_{\rm C}(t)$ denotes the total carbon mass located outside the black hole defined by
\begin{equation}
M_{\rm C}(t) \equiv \int_{V'} \rho_{0*} Z_{\rm C} dV.
\end{equation}
Figure~\ref{fig:A13} shows that $\langle Z_{\rm C} \rangle$ increases exponentially with time just before the formation of a black hole, but most of them are absorbed into the formed black hole. For both models A1 and A3, the value of $\langle Z_{\rm C} \rangle$ at $t-t_{\rm BH}=2000$ s is at most several tens of times larger than the initial values and mass fractions of protons and heliums are approximately constant throughout the collapse. 
Figure \ref{fig:A24} shows that several tens percent of heliums are burned into carbons just before the black-hole formation, but again almost all of them are absorbed into the formed black hole and the torus is composed primarily of heliums. 

Figures~\ref{fig:A13} and \ref{fig:A24} show that the nuclear burning more strongly occurs for slowly rotating models A1 and A2 than for rapidly rotating models A3 and A4. 
This fact can be understood in the following manner.
First, if a SMS core is rapidly rotating, the mass for the SMS core to become unstable to the gravitational collapse is heavier than the slowly rotating models because rotation strongly stabilizes the SMS core against gravitational collapse.
We note that the density of the collapsing core just before the black hole formation is smaller for the larger mass model 
because the length scale of the system is proportional to the mass of the black hole, $M_{\rm BH}$. 
Since the gravitational collapse proceeds approximately adiabatically, the temperature is also an increasing function of density. 
Hence for the lighter SMS core, the density and temperature at the moment of the black-hole formation are higher than for the heavier SMS core, and hence, the lighter SMS core induces stronger nuclear burning.

Figure~\ref{fig:nuc14} displays the time evolution of the total energy generation rate of the nuclear burning, $\dot{E}_{\rm nuc}$, defined by
\begin{equation}
\dot{E}_{\rm nuc}(t) \equiv \int_{V'} \rho_{0*} (\dot{q}_{\rm CNO} + \dot{q}_{3\alpha}) dV.
\end{equation}
This rate exponentially increases just before the black-hole formation. 
We also find that for models A1 and A3, $\dot{E}_{\rm nuc}$ is dominated by CNO cycle both before and after the collapse, and only just before the collapse, the energy generation rate by CNO cycle and triple-alpha reactions are comparable. 

Figure~\ref{fig:neu14} displays the time evolution of the total generation rate of neutrinos, $\dot{E}_\nu$, defined by
\begin{equation}
\dot{E}_{\nu}(t) \equiv \int_V \rho_{0*} (\dot{q}_{\rm photo} + \dot{q}_{\rm pair} + \dot{q}_{\rm plasma}) dV,
\end{equation}
where $\dot{q}_{\rm photo},~\dot{q}_{\rm pair}$, and $\dot{q}_{\rm plasma}$ are the neutrino generation rate  resulting from the photo neutrino process, pair neutrino process, and plasma neutrino process, respectively. 
We note again that the effect of neutrino cooling is not included in our simulations.
We remark that the neutrino luminosity of the collapsing SMS core would be much smaller than $\dot{E}_{\nu}$ because most of the generated neutrinos would be absorbed by the black hole. 

In the early phase of the collapse, $\dot{E}_{\nu}$ is dominated by photo-neutrino emission for all the models. Just before the black-hole formation, $\dot{E}_\nu$ is dominated by pair-neutrino emission for all the models. After the collapse, $\dot{E}_\nu$ is dominated by photo-neutrino emission for model A1 and by pair-neutrino emission for models A2--A4. 

$\dot{E}_{\rm nuc}$ and $\dot{E}_{\nu}$ in Table~\ref{t-3} show the values for models A1--A4 at $t-t_{\rm BH}=2000$ s.  
These values approximately denote the nuclear burning and neutrino cooling rates of the torus.
We find that for both $\dot{E}_{\rm nuc}$ and $\dot{E}_{\nu}$, rapidly rotating models A3 and A4 have larger values than the slowly rotating models A1 and A2, respectively. 
In addition, helium-burning models A2 and A4 have larger rates than hydrogen-burning models A1 and A3:
The former is due to the fact that for the rapidly rotating models, their tori are more massive than the slowly rotating models as discussed above.  The latter is due to the fact that for the helium-buring models, their density and temperature are higher than for the hydrogen-burning models. We list the maximum density, $\rho_{\rm max}$, and temperature, $T_{\rm max}$, of the torus  at $t-t_{\rm BH}=2000$ s for models A1--A4 in Table~\ref{t-3}. The maximum temperature and density for helium-burning models A2 and A4 are approximately 2 and 20 times larger than for hydrogen burning models A1 and A3, respectively.

In Table~\ref{t-3}, we also list the internal energy of the torus, $E_{\rm tor}$, heating timescale of the torus by nuclear burning, $\tau_{\rm nuc}$,  and rotation period at the density maximum of the torus, $\tau_{\rm dyn}$ at $t-t_{\rm BH} = {\rm 2000}$ s for models A1--A4, respectively. 
Here, $E_{\rm tor}$ is defined by
\begin{equation}
E_{\rm tor}(t) \equiv \int_{V'} \rho_{*} \epsilon dV,
\end{equation}
where $\rho_{*} \equiv \rho \sqrt{-g}c u^t$.
We calculate $\tau_{\rm nuc}$ by $\tau_{\rm nuc}= E_{\rm tor}/\dot{E}_{\rm nuc}$. 
$\tau_{\rm nuc}$ for models A1--A4 is approximately $1\times 10^{10}~{\rm s},~9\times 10^7~{\rm s},~7\times 10^{10}~{\rm s}$,and $2\times 10^{8}$ s, respectively. 
Hence, the nuclear reaction timescales is much longer than the dynamical timescale of the torus.

Before closing this section, we consider the possible effect of viscosity, which is not taken into account in our present study but it could play an important role for the evolution of the torus surrounding the black hole in reality.
In the $\alpha$-viscous model \cite{1973A&A....24..337S}, the viscous heating timescale is written approximately as 
\begin{equation}
\tau_{\rm vis} \sim \alpha^{-1}\tau_{\rm dyn}.
\label{s-21}
\end{equation}
With the values of $\tau_{\rm dyn}$ listed in Table~\ref{t-3}, $\tau_{\rm vis}$ is $110\alpha^{-1}~{\rm s},~25\alpha^{-1}~{\rm s},~190\alpha^{-1}~{\rm s}$, and $46\alpha^{-1}$ s for models A1--A4, respectively.
This suggests that  the viscous heating timescale would be much shorter than the nuclear heating timescale unless $\alpha < 10^{-6}$ for all the models. 
Thus if we want to predict the long-term evolution of the torus, the viscous heating would be necessary. 

\subsection{Effects of nuclear burning}
\label{effect}
The SMS cores do not explode for all the models. 
{For comparison, we also perform simulations for the same initial conditions as models A1--A4 but putting out nuclear burning.  
We find that the difference of the values of $M_{\rm BH},~a_{\rm BH},~E_{\rm tor},~T_{\rm max}$, and $\rho_{\rm max}$ between these models and models A1--A4  are less than $1\%$.} In this section, we clarify the reason why the nuclear burning does not play an important role during the collapse.  

\begin{figure}[htbp]
	\includegraphics[angle=0, width=1\linewidth]{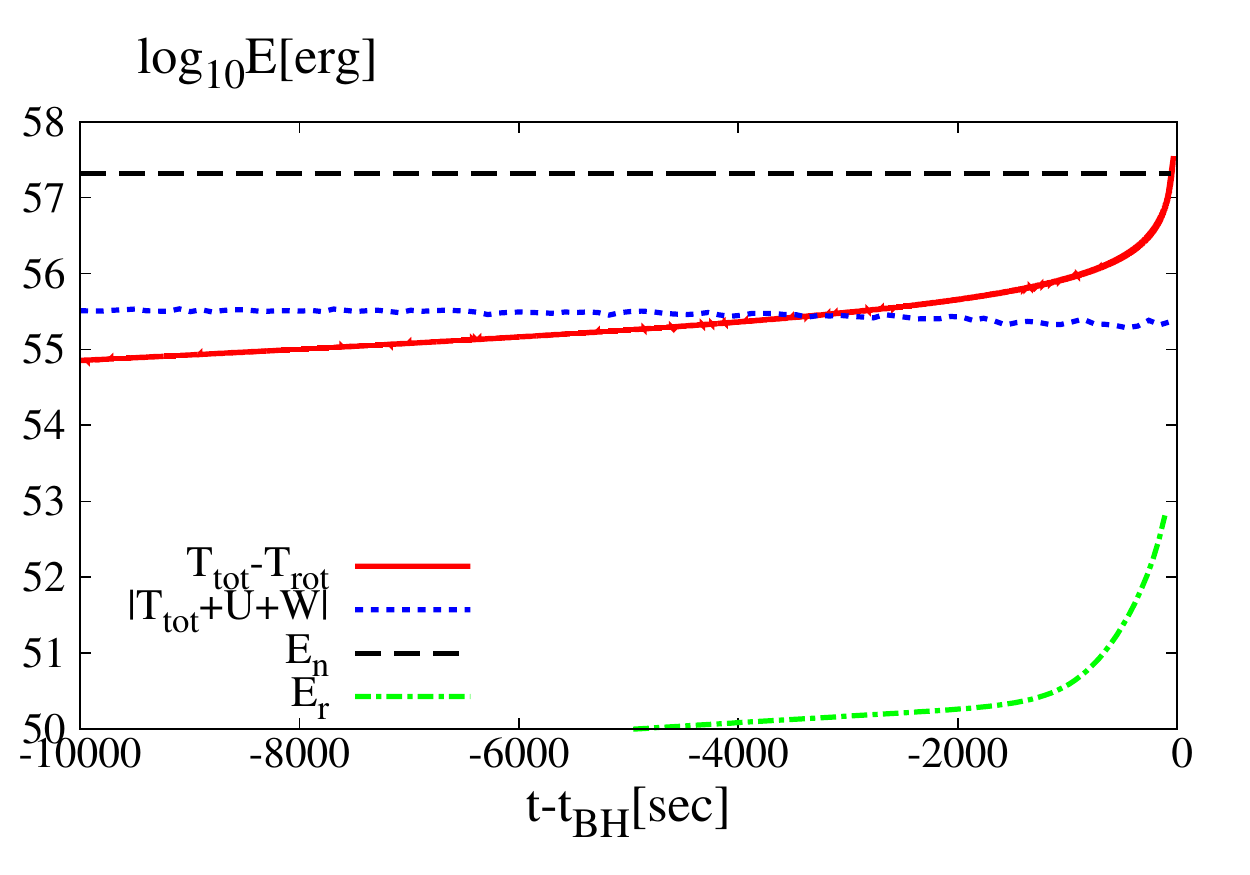}
	\caption{Time evolution of several key energies for model A1. $T_{\rm tot},~T_{\rm rot},~U,~W,~E_{\rm n}$, and $E_{\rm r}$  denote the total kinetic energy, the rotational kinetic energy, the total internal energy, the total gravitational energy, the total rest-mass energy which could be released by nuclear burning, and the total released rest-mass energy by nuclear burning, respectively. We plot $T_{\rm tot}-T_{\rm rot}$ (red-solid), $|T_{\rm tot}+U+W|$ (blue-dotted), $E_{\rm n}$ (black-dashed), and $E_{\rm r}$ (green-dashed-dotted), respectively.} 
	\label{fig:gravA1}
\end{figure}
First, we pay attention to the evolution of various energies of the gravitational collapse.
Figure~\ref{fig:gravA1} plots the evolution of the total kinetic energy, $T_{\rm tot}$, the rotational kinetic energy, $T_{\rm rot}$, the total internal energy, $U$, the total gravitational energy, $W$, the total rest-mass energy which could be released by nuclear burning, $E_{\rm n}$, and the total released rest-mass energy by nuclear burning, $E_{\rm r}$, for model A1, respectively. Here, $T_{\rm tot}$, $T_{\rm rot}$, $U$, and $W$ are defined by
\begin{eqnarray}
T_{\rm tot}(t) &\equiv& \int_V  \frac{1}{2} \rho_{0*} c^2 (1-(\alpha c u^t)^2) dV,\\
T_{\rm rot}(t) &\equiv&  \int_V \frac{1}{2c} \rho_{0*} h_0v^\varphi u_\varphi dV,\\ 
U(t) &\equiv& \int_V \rho_{*} \epsilon dV,
\end{eqnarray}
\begin{equation}
W(t)\equiv M_{\rm ADM}(t)c^2 - M_{\rm total}(t)c^2-T_{\rm tot}(t) - U(t), 
\end{equation} 
where $M_{\rm ADM}$ is the ADM mass (gravitational mass) defined by
\begin{eqnarray}
M_{\rm ADM}(t) &\equiv& \frac{c^2}{16\pi G}\int_V \left[ - \tilde{R} + \frac{16 \pi G}{c^4}\rho_{\rm h} \psi^5 \right.  \nonumber \\ 
&+&\left.  \psi^{-7} \bar{A}^{ij}\bar{A}_{ij}  -\frac{2}{3} \psi^5 K^2  \right] dV.
\end{eqnarray}
 Here, $\psi = \rho_{\rm g}^{-1/2}$, $\bar{A}^i_j \equiv \psi^6 \tilde{A}^i_j$, $\rho_{\rm h} \equiv \rho h (\alpha c u^t)^2 - P$, and $\tilde{R}$ is the Ricci scalar with respect to $\tilde{\gamma}_{ij}$, respectively. The definition of $T_{\rm rot}$ is equivalent to  Eq.~(\ref{ktots}). $T_{\rm tot}-T_{\rm rot}$ approximately denotes the kinetic energy associated with the infalling motion. $E_{\rm n}$ and $E_{\rm r}$ are defined by
\begin{eqnarray}
E_{\rm n} (t)&\equiv& \left (   \frac{  m_{\rm C} }{12   }-m_{\rm p}-\frac{m_{\rm e}}{2}  \right ) \frac{M_{\rm total}(t) c^2}{m_{\rm p}}Y_{\rm p}(t) \nonumber \\
&+&\left (   \frac{  m_{\rm C} }{3}-m_{\alpha}\right ) \frac{M_{\rm total}(t) c^2}{m_{\alpha}}X_{\rm \alpha}(t), \label{s-21.1}\\
E_{\rm r} (t)&\equiv& E_{\rm n}(0)-E_{\rm n}(t), \label{s-21.2}
\end{eqnarray}
respectively.
We find that until the black-hole formation, $E_{\rm r} \ll T_{\rm tot}-T_{\rm rot}$ is satisfied. This fact shows that the nuclear burning cannot halt the collapse of the infalling matter. 
It is also found that $T_{\rm tot}-T_{\rm rot}>E_{\rm n}$ is satisfied at $t-t_{\rm BH}\approx 0$ s. 
This fact indicates that after this time, nuclear reactions would become unable to induce explosion. 
We note that for models A2—A4, the relations $E_{\rm r} \ll T_{\rm tot}-T_{\rm rot}$ and $T_{\rm tot}-T_{\rm rot} > E_{\rm n}$ for $t-t_{\rm BH} \approx 0$ are satisfied. Thus, for any models employed in this paper, the collapse to black holes cannot be halted by the nuclear burning.

Next, we consider the property of the torus. As described in Sec.~\ref{nuclear}, the nuclear heating timescale is $\gtrsim 10^8$ s for all the models employed in this paper. By contrast, we find that the timescale of the formation of the torus is $\lesssim 10^4$ s for all the models. This is much shorter than the nuclear heating timescale. Hence the nuclear burning cannot modify the property of the torus during the gravitational collapse.

Finally, we consider the property of the outflow. The outflow occurs as a result of the shock heating on the surface of the torus. 
However, the density and temperature in the shock are not very high. 
Hence the nuclear burning also cannot affect the outflow. We describe the formation process of the outflow in detail in Sec.~\ref{outflow}.
	\subsection{Torus mass}
\begin{figure}[htbp]
		\includegraphics[angle=0, width=1\linewidth]{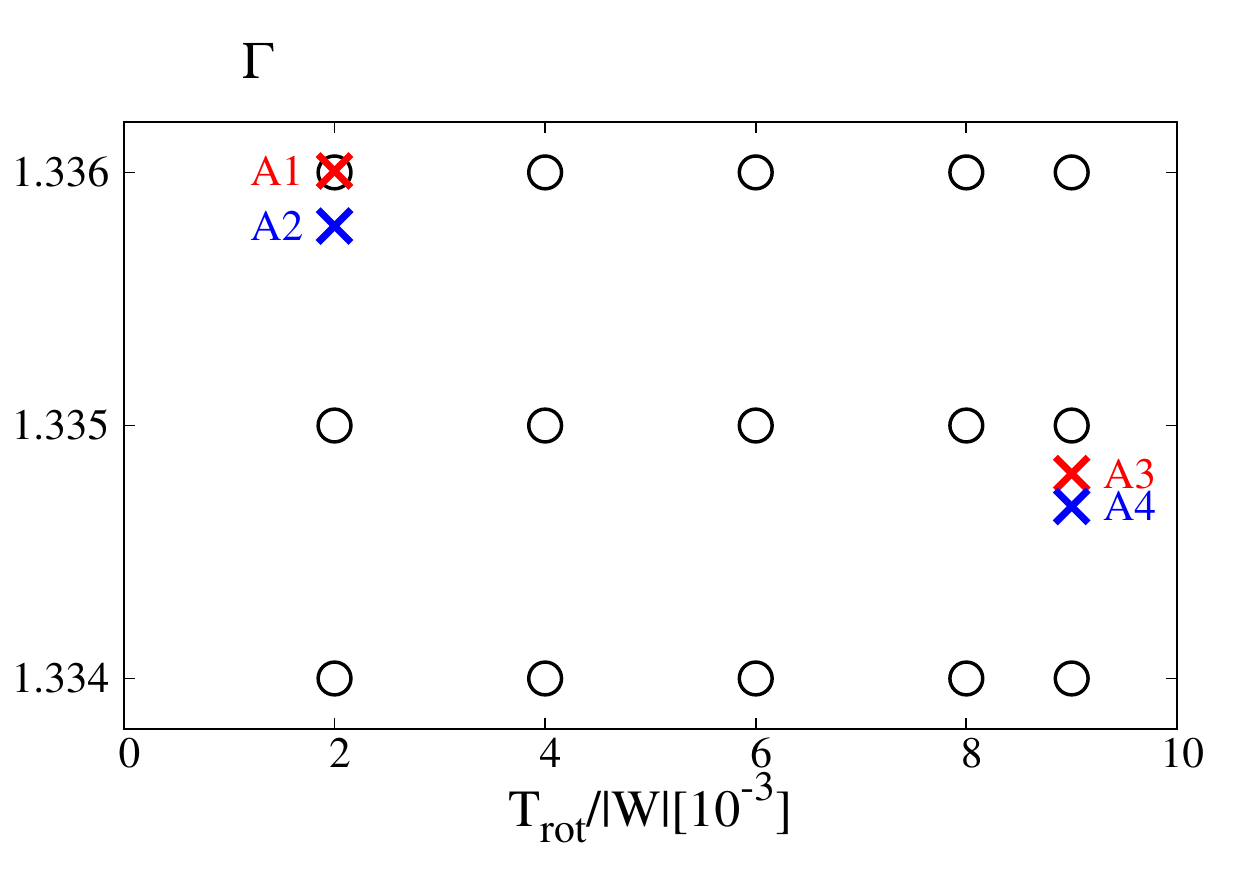}
		\caption{Initial values of $\Gamma$ and $T_{\rm rot}/|W|$ for the hydrogen-burning models (red crosses), the helium-burning models (blue crosses), and  the N models~(open circles). The crosses correspond to models A1--A4.}
		\label{fig:gamma}
	\end{figure}

	As we found in the previous section, the effect of nuclear burning only weakly affects the collapse dynamics and the properties of the torus and outflow. Then we search for the relation between $M_{\rm ter}$ and $T_{\rm rot}/|W|$ for given adiabatic constants neglecting nuclear burning. 
	For this purpose, we performed simulations for additional initial conditions (hereafter referred to as N models). 
	The initial conditions are chosen taking into account the following facts: 
	The red and blue crosses in Fig.~\ref{fig:gamma} show that the values of $\Gamma$ and $T_{\rm rot}/|W|$ for models A1--A4 have the relation of 
    $\Gamma=1.334$--$1.336$ for $T_{\rm rot}/|W|=0.002$--$0.009$. 
	Then we uniformly select 15 initial conditions 
	including 3 different adiabatic constants $\Gamma =1.334,~1.335$ and $1.336$ and 5 different rotation parameters $T_{\rm rot}/|W| = 0.002, 0.004, 0.006, 0.008$, and $\approx 0.009$ with the same EOS as Eq.~(\ref{s-9}). The simulation is performed with $D_{\rm T}=0.5$. 
	As we already found, the properties of the formed black hole and the mass of the torus depend only weakly on the value of $D_{\rm T}$.  
	The open circles in Fig.~\ref{fig:gamma} denote these models.
	
\begin{figure}[htbp]
	\includegraphics[width=1\linewidth]{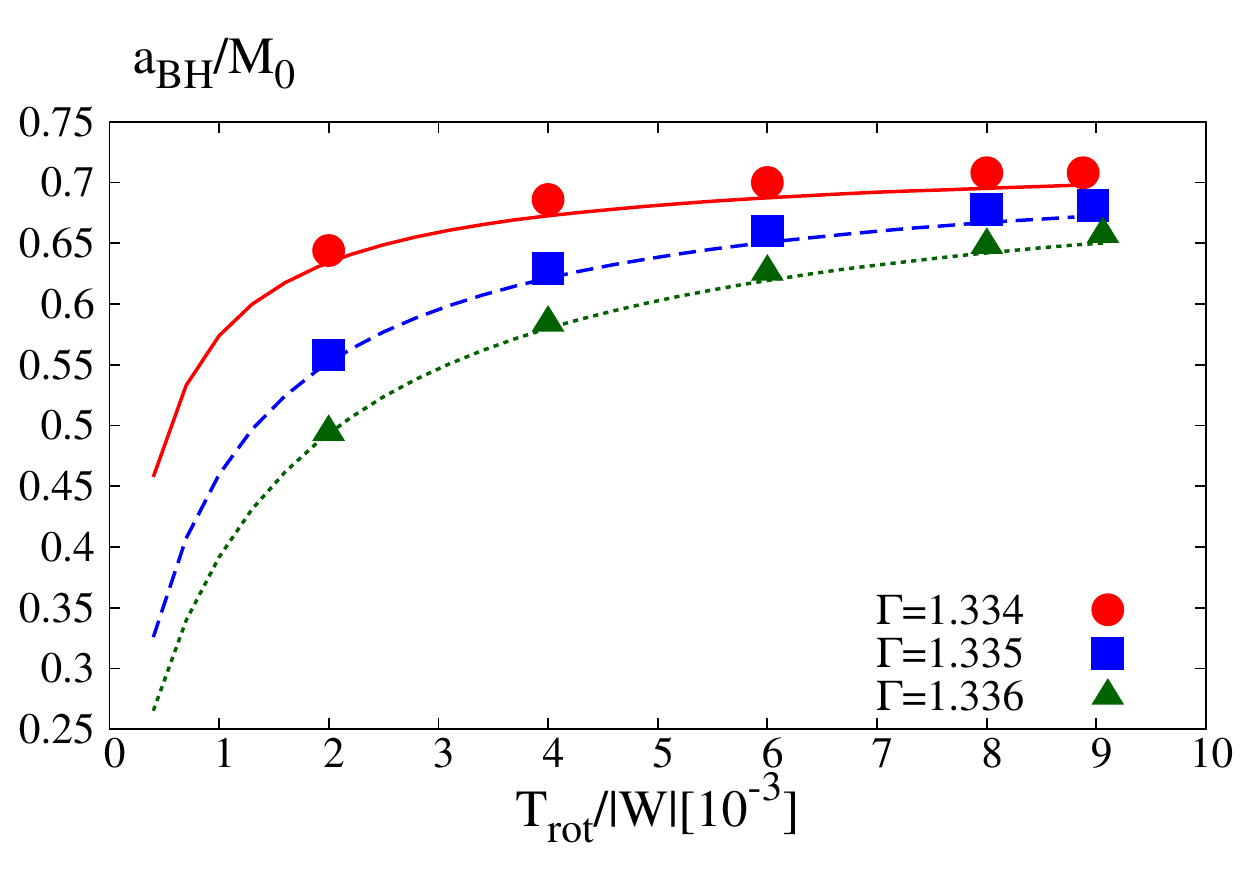}
	\caption{The dimensionless spin parameter of the remnant black hole as a function of the rotation parameter $T_{\rm rot}/|W|$ for three values of the  adiabatic constant $\Gamma$. The filled circles, squares, and triangles are the results of the numerical simulation and the solid, dashed, and dotted curves are the analytical predictions of~Ref.~\cite{0004-637X-818-2-157} for $\Gamma= 1.334,~1.335$, and $1.336$, respectively. }
	\label{fig:ah}
	\end{figure}

\begin{figure}[htbp]
		\includegraphics[width=1\linewidth]{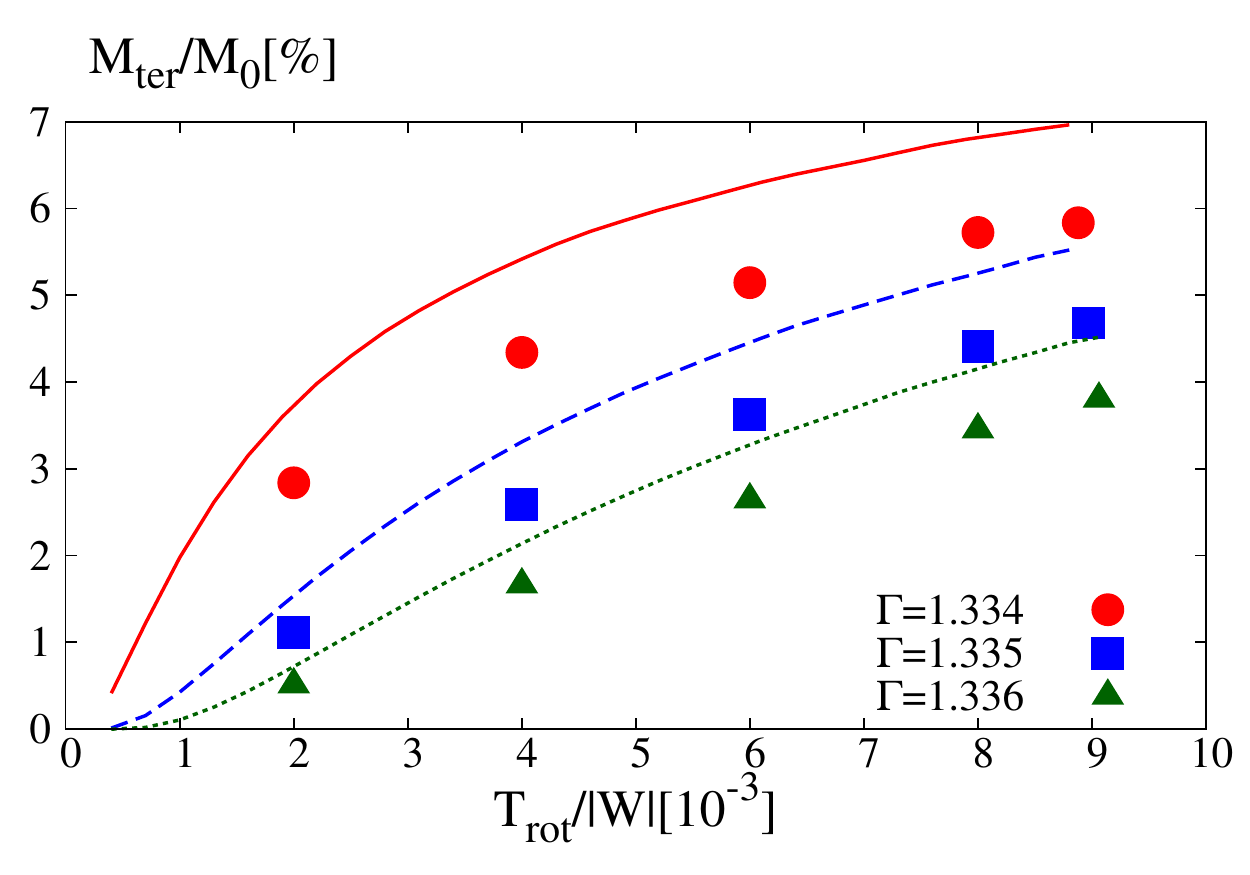}
		\caption{$M_{\rm ter}$ as a function of $T_{\rm rot}/|W|$ for three values of adiabatic constant $\Gamma$. The filled circles, squares, and triangles are the results of the numerical simulation and the solid, dashed, and dotted curves are the analytical predictions of~Ref.~\cite{0004-637X-818-2-157} for $\Gamma= 1.334,~1.335$, and $1.336$, respectively. }
		\label{fig:rem}
	\end{figure}

	We define the mass of the final state of the torus by
	\begin{equation}
	M_{\rm tor}\equiv M_{\rm ter} - M_{\rm eje}.
	\end{equation} 
	As we found in Sec.~\ref{over}, $M_{\rm eje}$ is much smaller than $M_{\rm ter}$, and then, $M_{\rm tor}$ can be approximated by $M_{\rm ter}$.
	Figures~\ref{fig:ah} and \ref{fig:rem} depict the dimensionless spin parameter of the remnant black holes and $M_{\rm ter}$, respectively. 
    The filled circles, squares, and triangles are the results of the numerical simulation and the solid, dashed, and dotted curves denote the relations obtained by the analytical predictions of~Ref.~\cite{0004-637X-818-2-157} for $\Gamma= 1.334,~1.335$, and $1.336$, respectively.
	
	In Ref.~\cite{0004-637X-818-2-157}, we inferred the mass and spin of the remnant black holes in the following manner. 
	First, we employed three assumptions that (i) the collapse proceeds  in an axisymmetric manner, (ii) the angular momentum transport due to the viscosity is negligible, and 
	(iii) nuclear burning does not halt the collapse. The first and second assumptions are the same as those we employed in the simulations. 
	The third assumption is shown to be appropriate in Sec.~\ref{effect}. 
	
	We also assumed that the gravitational collapse proceeds in the following manner. 
	First, a seed black hole is formed at the center of the collapsing SMS core 
	and it dynamically grows while sequentially absorbing fluid elements from lower values of specific angular momentum, $j$, defined by 
	\begin{equation}
	j \equiv h_{\rm 0} u_{\rm \varphi}.
	\end{equation} 
	We then calculated the mass and angular momentum of the hypothetically growing black hole at each moment by calculating Eqs.~(37) and (38) of Ref.~\cite{0004-637X-818-2-157} and subsequently got the dimensionless spin parameter of the black hole. 
	
	Then, assuming that the black hole is Kerr black hole, we can calculate $j_{\rm ISCO}$ by using Eq.~(2.21) of Ref.~\cite{1972ApJ...178..347B}. 
	Here $j_{\rm ISCO}$ is the specific angular momentum, which is needed for a test particle to rotate at an innermost stable circular orbit~(ISCO) in the equatorial plane around the black hole. 
	We assumed that the growth of the black hole would terminate at the moment at which $j$ becomes larger than $j_{\rm ISCO}$.
	In this assumption, we neglect the pressure and geometry of the torus because $j_{\rm ISCO}$ is determined by calculating the geodesic equation of test particles moving in the equatorial plane. 
	
	Figure~\ref{fig:ah} shows that the dimensionless spin parameter matches well with the prediction, but Fig.~\ref{fig:rem} shows that $M_{\rm ter}$ is slightly overestimated in the analytic prediction. This is likely due to the fact that each fluid element of the SMS core falls toward the black hole with an elliptical orbit, and then, some fluid elements fall into the black hole even when their specific angular momentum is larger than $j_{\rm ISCO}$.  
	The pressure and geometry of the torus may also affect the result. Nevertheless, the analytic calculation predicts the mass of the torus within  $30\%$ error. 
	
	We conclude that if a SMS core is rotating with $T_{\rm rot}/|W| \gtrsim 0.002$, (i) a few percent of the initial mass forms a torus surrounding the central black hole irrespective of the nuclear burning phases of SMS cores, and (ii) the dimensionless spin of the remnant black hole is in the range between $\approx 0.5$ and $0.7$. 
	

\section{outflow}
\label{outflow}
As found in Sec.~\ref{result}, a fraction of the mass of the SMS core is ejected soon after the black hole is formed. In this section, we describe the properties of the outflow in detail. For this analysis, we define several quantities of the outflow in Appendix~\ref{app2}. 
\subsection{Formation process of the outflow}
\label{formation}
\begin{figure*}[htbp]
	\includegraphics[scale=0.65, trim=30 80 0 50, clip]{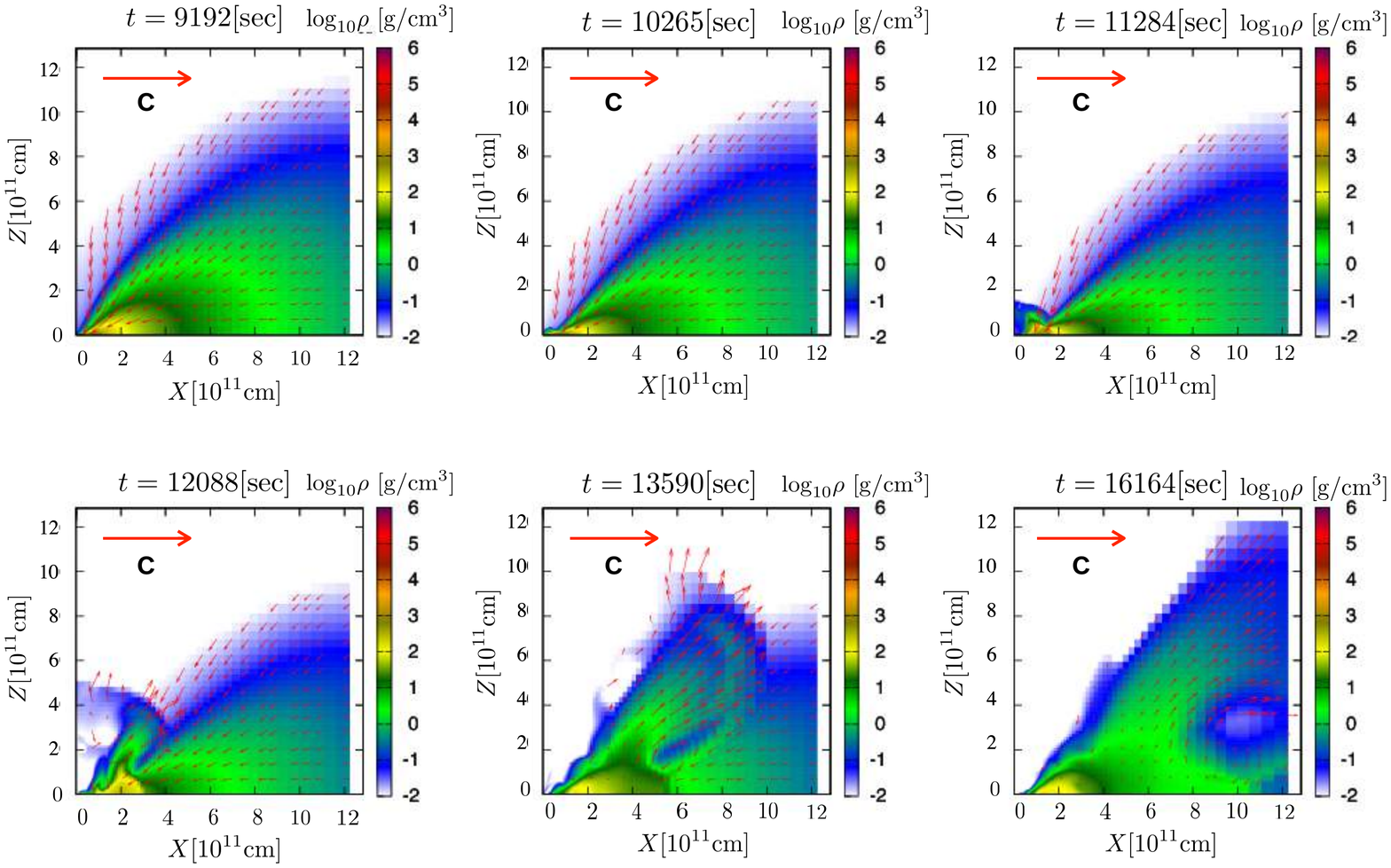}
	\caption{The density profiles near the formation time of the torus surrounding the black hole. They show zoom-in views of the central region. The red arrows denote the velocity vectors and their length scale is normalized as indicated in the upper left-hand corner of each snapshot. Just after the torus formation, a strong shock and a resulting bubble are formed and a fraction of the torus matter is ejected.  }
	\label{fig:shock}
\end{figure*}

First, we describe the mechanism to drive the outflow. Figure~\ref{fig:shock} displays the snapshots of the density profile at the formation of the torus for the N model with $\Gamma=1.336$ and $T_{\rm rot}/|W|\approx 0.009$. For this model, the chemical composition is set to be the same as the hydrogen-burning models.

The 1st panel shows the density profile at the launch time of the outflow. The fluid elements conserve their specific angular momentum, and hence, if their initial specific angular momentum is slightly larger than $j_{\rm ISCO}$ for the formed black hole, they fall toward the vicinity of the black hole due to the centrifugal force barrier. At the same time, a fraction of the fluid elements in the torus is pushed inward by the inertia of the torus which has small infall velocity. Then, the material falling from a high latitude toward the inner edge of the torus hits the inner part of the torus, and, due to the strong encounter among the fluid elements, shocks are formed. As a result, a dense bubble is formed by the shock heating at the very central region, i.e., $X<1 \times 10^{11}$ cm~(2nd panel). 

The formed bubble immediately expands vertically because the density above the torus is very small~(3rd panel). This bubble feels strong centrifugal force, and hence, it moves toward the surface of the torus. Then most of the fluid components in the bubble is re-absorbed by the torus. However, a fraction of them which expands vertically can avoid colliding with the torus, and then, it spreads outward~(4th--6th panels). This situation is similar to the hot bubble formation in a collapsar model \cite{2006ApJ...641..961L}. 
The formation process of the outflow is also discussed in Ref.~\cite{PhysRevD.76.084017}.

\subsection{Properties of the outflow}
\label{time}
\begin{figure}[htbp]
	\includegraphics[width=\linewidth]{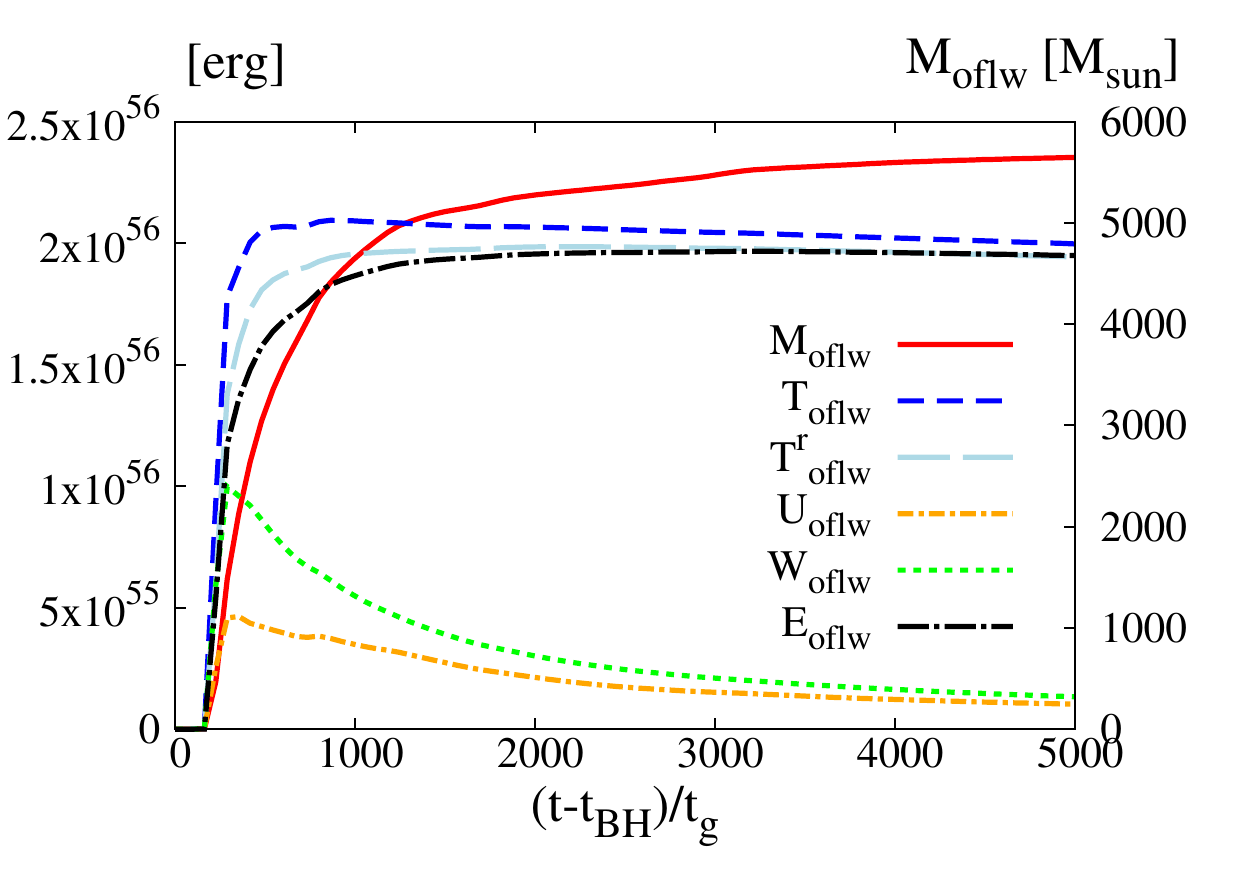}
	\caption{The time evolution of several key quantities of the outflow for model A3. We plot the total mass of the outflow (red-solid), and the total kinetic energy~(blue-dashed), the radial component of the total kinetic energy~(light blue-long-dashed), the total internal energy~(orange-dashed-dotted), the total gravitational potential energy~(green-dotted), and the total energy~(black-long-dashed-dotted), respectively.}
	\label{fig:outmA3}
\end{figure}

Next, we summarize the property of the outflow. Figure  \ref{fig:outmA3} shows the time evolution of several key quantities of the outflow for hydrogen-burning model A3 with $D_{\rm T}=0.5$ (note the dependence of the quantities of the outflow on $D_{\rm T}$ is discussed in Appendix \ref{initp}). 
We plot the total mass $M_{\rm oflw}$ (red-solid), and the total kinetic energy $T_{\rm oflw}$~(blue-dashed), the radial component of the total kinetic energy~$T_{\rm oflw}^r$~(light blue-long-dashed), the total internal energy~$U_{\rm oflw}$ (orange-dashed-dotted), the total gravitational potential energy~$|W_{\rm oflw}|$ (green-dotted), and the total energy~$E_{\rm oflw}$ (black-long-dashed-dotted), respectively. They are defined by Eqs.~(\ref{total}), (\ref{ttot}), (\ref{trtot}), (\ref{utot}), and  (\ref{wtot}), respectively. 
Here, $t_{\rm g}$ is the light crossing time defined by
\begin{equation}
t_{\rm g} \equiv \frac{GM_{0}}{c^3},
\label{tg}
\end{equation}
where $M_0$ is the initial mass of the SMS core. 

For $t-t_{\rm BH}\lesssim 1000t_{\rm g}$, the mass and total energy are increasing because the outflow propagates while collecting mass of the SMS core. 
At $t-t_{\rm BH}\approx 1000t_{\rm g}$, 
most of the outflow matter reaches the surface of the SMS core, and hence, the increase of the mass and total energy becomes slower.
The final value of the outflow mass is $\approx 5700M_\odot$ in this model. 

For $t-t_{\rm BH}\gtrsim 1000t_{\rm g}$, the total kinetic energy starts decreasing slightly.
We speculate that this degrease is due to the following reasons: 
(i) the kinetic energy of the fluid elements decreases when moving in the gravitational potential of the central object, 
(ii) the shock occurred in the outflow dissipates the kinetic energy of the outflow, and thus, a fraction of the outflow becomes bound again,
and (iii) a numerical error induces the decrease of the total energy.
The total energy at $t-t_{\rm BH}=5000t_{\rm g}$ is approximately $1.8\times 10^{56}$ erg. 
The possible error size to this would be of order $10^{54}$ erg as discussed above.

The amount of total kinetic energy is approximately 10 times larger than rotational kinetic energy, internal energy, and gravitational potential energy. 
Hence the radial component accounts for most of the total energy. 
Hence the outflow would propagate through the envelope of the SMS approximately radially. 
We find that these  properties are satisfied for all the models. 

The reason why  the properties of the outflow do not depend on the chemical composition of the SMS core can be understood in the following manner.
As described in Sec.~\ref{effect}, the effect of nuclear burning can be neglected throughout the collapse. 
If we neglect the effect of nuclear burning, the gravitational collapse proceed approximately adiabatically before the formation of the shocks. 
Then, the EOS of the SMS core can be approximated by the polytropic EOS with $\Gamma \approx 4/3$  because the SMS core is radiation pressure dominated.
Thus if the same rotation parameter is taken, the dynamics of the gravitational collapse is qualitatively the same irrespective the chemical composition.
After the shock formation, the system becomes no longer adiabatic. However, qualitatively the same collapse may induce qualitatively the same outflow.
 
\begin{table}[htpb]
	\caption{Total mass and energy of the outflow for models A1--A4 with $D_{\rm T}=0.5$. $M_{\rm oflw}$ and $E_{\rm oflw}$ denote the total mass and energy defined by Eq.~(\ref{tot}) at $t-t_{\rm BH}=5000t_{\rm g}$, respectively. We note that $E_{\rm oflw}$ would have error of order $10^{54}$ erg as discussed in Sec.~\ref{time}.
	 $E_{\rm oflw}(D_{\rm T}=0)$ is the value of $E_{\rm oflw}$ at $t-t_{\rm BH}=5000t_{\rm g}$ with $D_{\rm T}=0$ for models A2 and A4 estimated in Appendix \ref{initp}. }
	\begin{tabular}{|c||c|c|c|} \hline
		Model &  $M_{\rm oflw}~(M_\odot)$ & $E_{\rm oflw}~({\rm erg})$&$E_{\rm oflw}(D_{\rm T}=0)~({\rm erg})$ \\ \hline \hline
		A1 &$ 320 $&$ 1.0 \times 10^{55} $&$~$\\ \
		A2 &$ 90 $&$ 2.8 \times 10^{54}$ &$2.8 \times 10^{54}$\\ \
		A3 &$ 5700 $&$ 1.8\times 10^{56} $&$~$\\ \
		A4 &$ 1400 $&$ 4.7 \times 10^{55}$&$3.1 \times 10^{55}$\\  \hline
	\end{tabular}
	\label{t-o}
\end{table} 

\begin{figure}[htbp]
	\includegraphics[width=\linewidth]{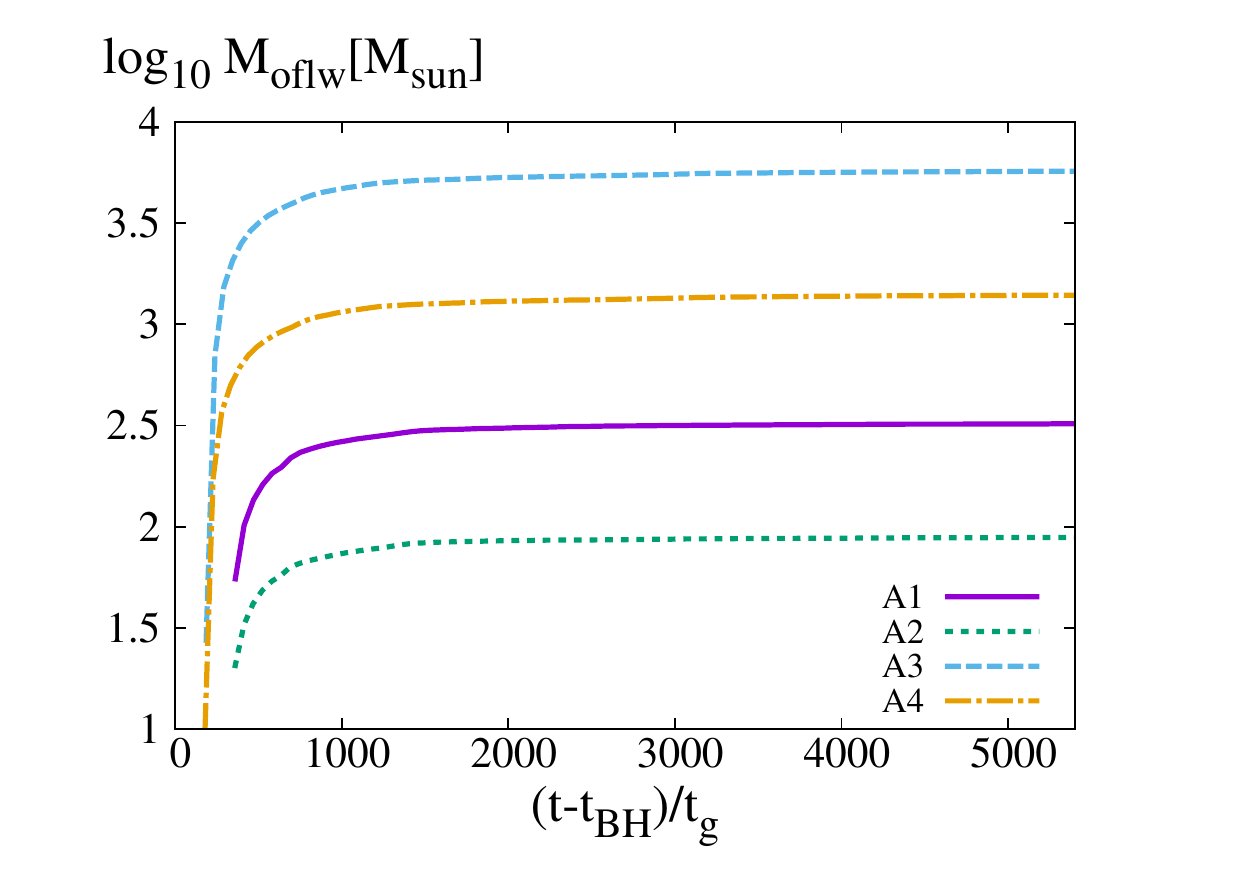}
	\caption{The time evolution of the total mass of the outflow, $M_{\rm oflw}$, for models A1 (purple-solid), A2 (green-dotted), A3 (light-blue-dashed), A4 (orange-dashed-dotted), respectively.}
	\label{fig:outmM}
\end{figure}

\begin{figure}[htbp]
	\includegraphics[width=\linewidth]{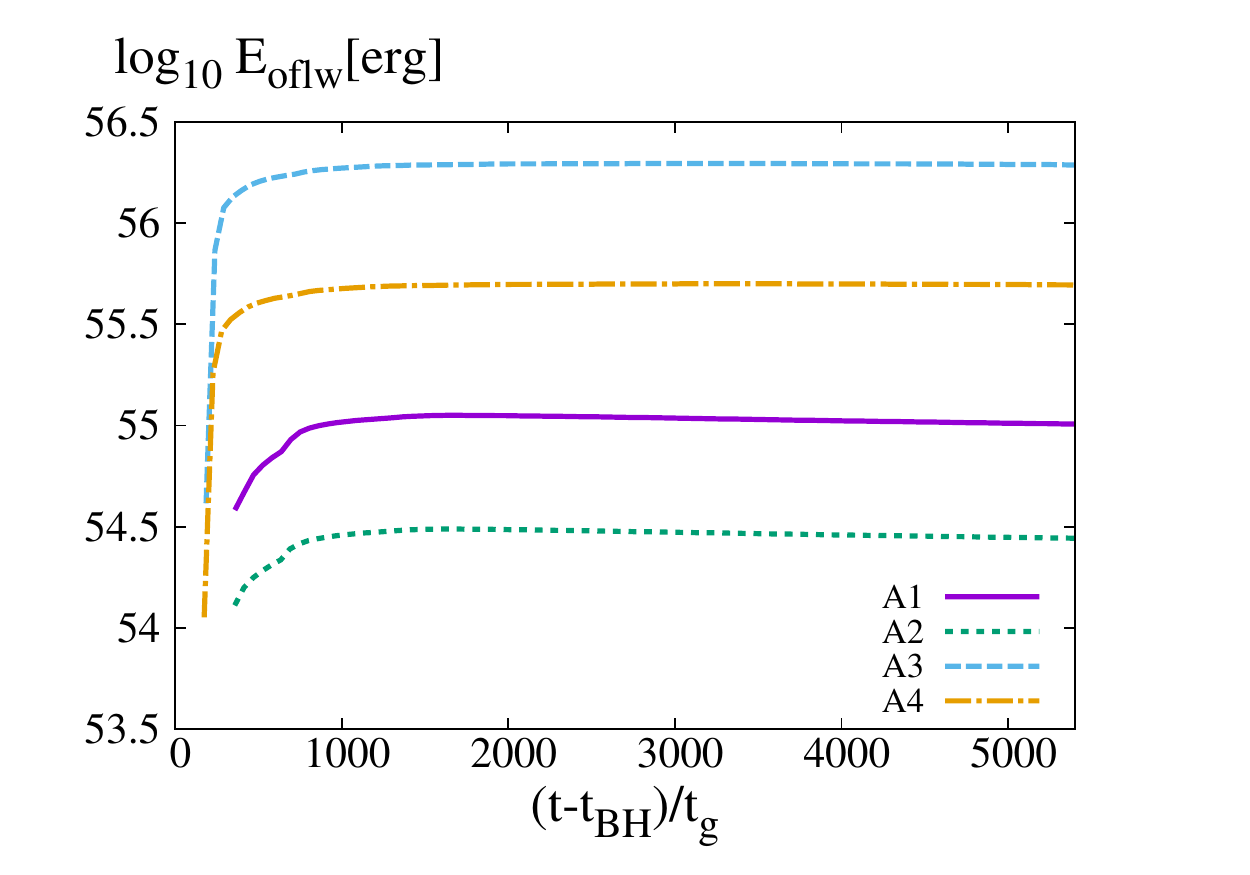}
	\caption{The time evolution of the total energy of the outflow, $E_{\rm oflw}$, for models A1 (purple-solid), A2 (green-dotted), A3 (light-blue-dashed), A4 (orange-dashed-dotted), respectively. We note that $E_{\rm oflw}$ would have error of order $10^{54}$ erg as discussed in Sec.~\ref{time}.}
	\label{fig:outmE}
\end{figure}

Figures~\ref{fig:outmM} and \ref{fig:outmE} denote the time evolution of the total mass and energy of the outflow for models A1--A4. 
We also list the total mass and energy of the outflow for models A1--A4 at $t-t_{\rm BH}=5000t_{\rm g}$ in Table~\ref{t-o}. 
For rapidly rotating models A3 and A4, the total mass and energy of the outflow is larger than for the slowly rotating models A1 and A2, respectively. 
This is due to the following reasons: (i) the initial mass of the SMS core for the rapidly rotating models are heavier than for the slowly rotating models, (ii) for the rapidly rotating models, each fluid element of the SMS core has larger specific angular momentum than for the slowly rotating models, and thus, more fraction of the mass escapes being swallowed by the black hole, and hence, can be ejected by the system due to centrifugal force barrier at the formation of the outflow as described in Sec.~\ref{formation}. 
\begin{figure}[htbp]
	\includegraphics[width=\linewidth]{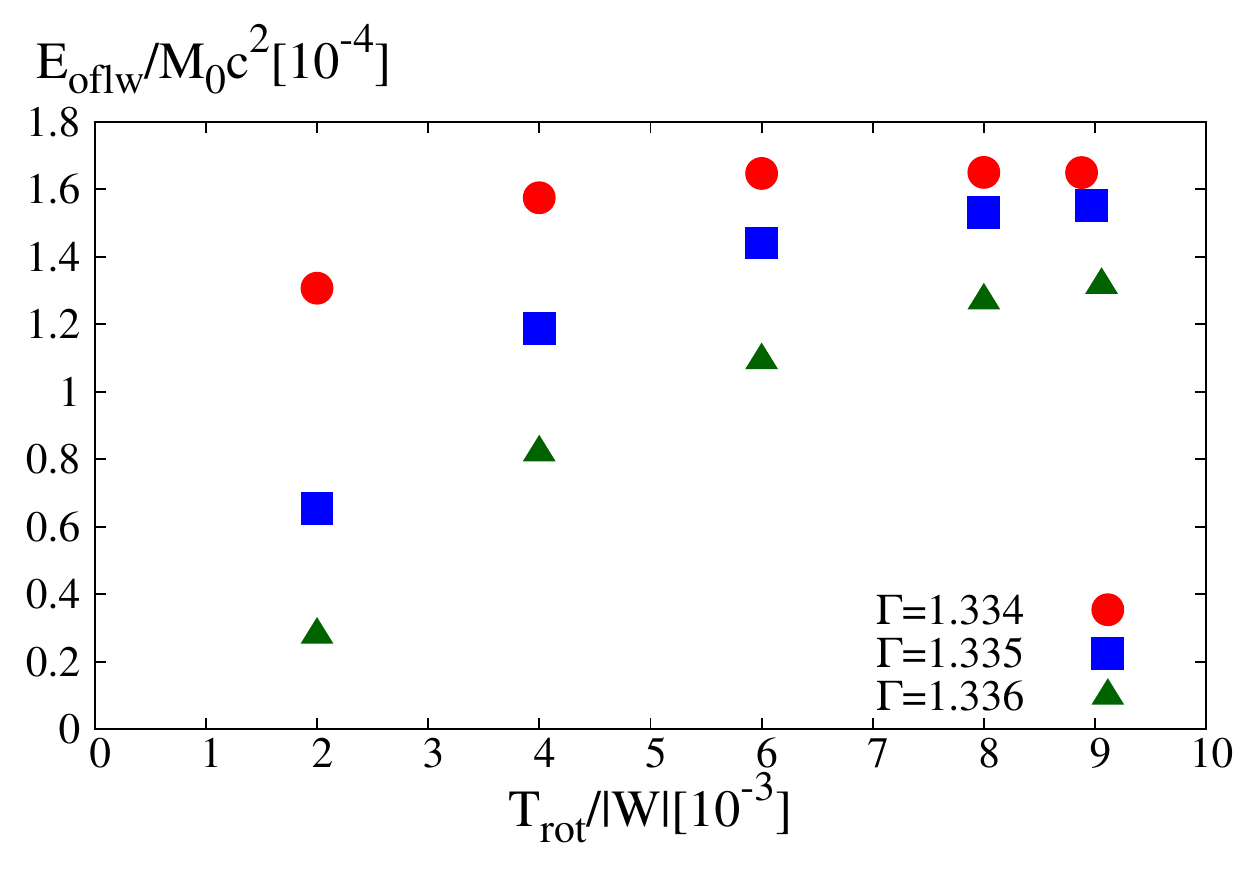}
	\caption{The relation between $E_{\rm oflw}$ and $T_{\rm rot}/|W|$ for the N models with $D_{\rm T}=0.5 $. The filled-circles, squares, and triangles denote the value for $\Gamma =1. 334,~1.335$, and $1. 336$ at $t-t_{\rm BH}=5000t_{\rm g}$, respectively. }
	\label{fig:TWgamma}
\end{figure}
\begin{figure}[htbp]
	\includegraphics[width=\linewidth]{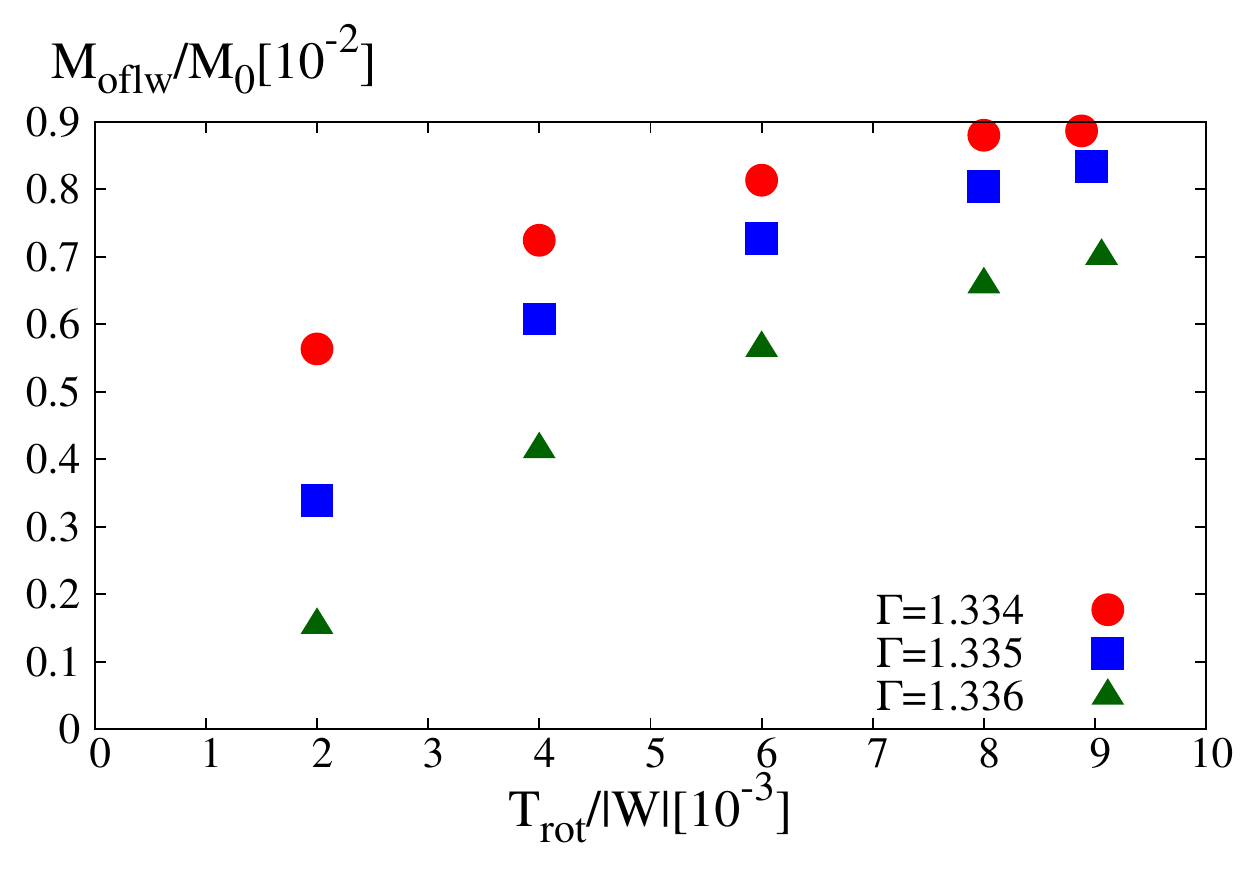}
	\caption{The relation between $M_{\rm oflw}$ and $T_{\rm rot}/|W|$ for the N models with $D_{\rm T}=0.5$. The filled-circles, squares, and triangles denote the value for $\Gamma =1. 334,~1.335$, and $1. 336$ at $t-t_{\rm BH}=5000t_{\rm g}$, respectively.. }
	\label{fig:TWmamma}
\end{figure}

Figures~\ref{fig:TWgamma} and \ref{fig:TWmamma} display the dependence of the mass and energy of the outflow on $T_{\rm rot}/|W|$ at $t- t_{\rm BH}=5000t_{\rm g}$ for N models. 
Here, we again choose $D_{\rm T}=0.5$. 
Both the mass and energy of the outflow increase as $T_{\rm rot}/|W|$ increases while the dependence is weaker for the ejecta energy.
For example, for $T_{\rm rot}/|W| \gtrsim 0.004$, the energy of the outflow is between $0.01\%$ and $0.018\%$ of the initial rest-mass energy.  
Thus the outflow energy for the SMS core collapse would be typically $\approx 10^{55} (M_0/10^5M_\odot)$ erg for a rapidly rotating model with $T_{\rm rot}/|W| \gtrsim 0.004$.


\begin{figure}[htbp]
	\includegraphics[width=\linewidth]{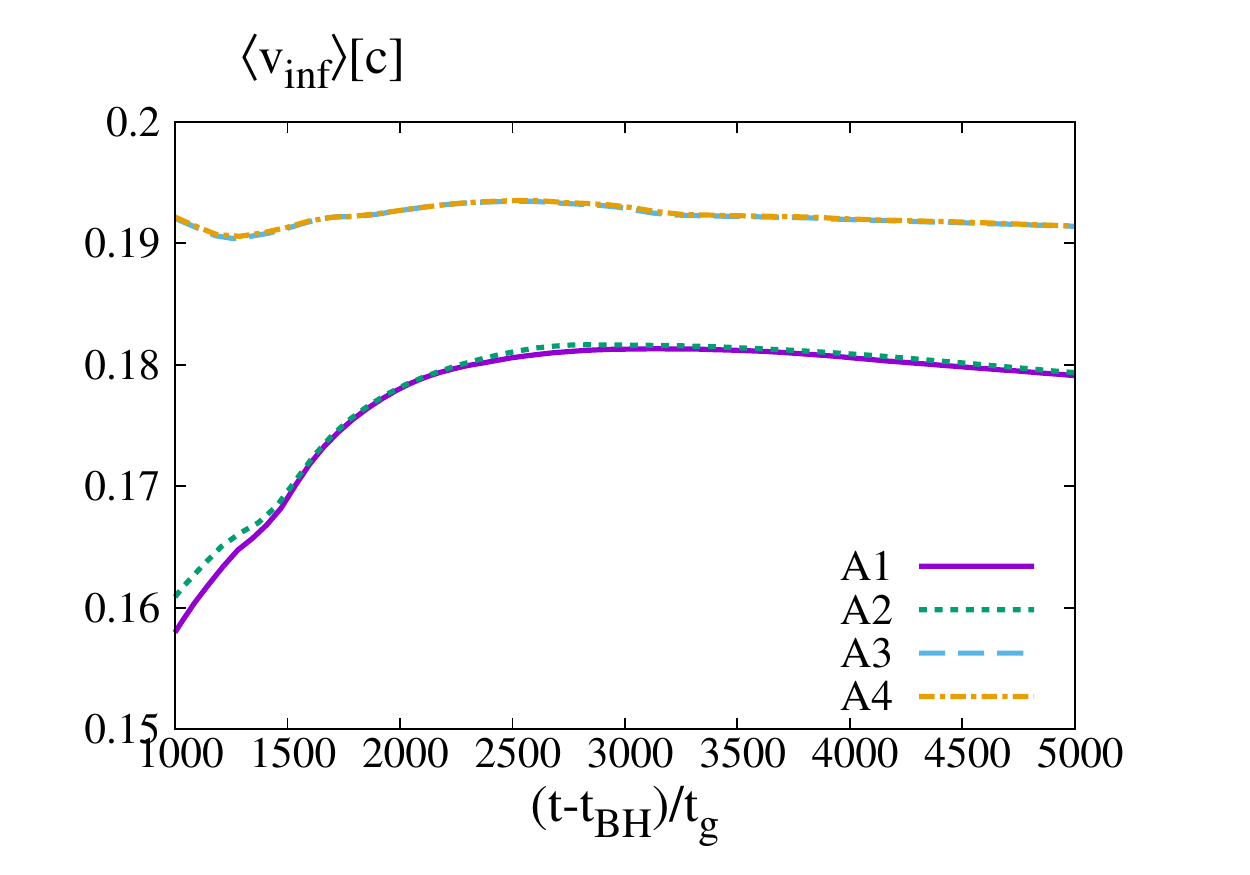}
	\caption{The time evolution of the average velocity of the outflow at infinity, $\langle v_{\rm inf} \rangle$, for models A1 (purple-solid), A2 (green-dotted), A3 (light-blue-dashed), A4 (orange-dashed-dotted), respectively.}
	\label{fig:vvvf}
\end{figure}

Figure~\ref{fig:vvvf} denotes the time evolution of the average velocity of the outflow at infinity, $\langle v_{\rm inf} \rangle$, defined by
\begin{equation}
\langle v_{\rm inf}(t) \rangle \equiv \sqrt{\frac{2(T_{\rm oflw}+W_{\rm oflw})}{M_{\rm oflw}}},
\label{vave}
\end{equation}
for models A1--A4.
At $t-t_{\rm BH}\gtrsim 3000t_{\rm g}$, $\langle v_{\rm inf} \rangle $ converges to $\approx 0.2c$ for all the models. 
We also find that $\langle v_{\rm inf} \rangle $ is $\approx 0.2c$ for all N models. 
Thus the outflow is escaped from the system with a subrelativistic velocity. 
In reality this outflow is likely to collide with the envelope of SMS surrounding its core region. It will be interesting to explore the phenomena associated with this collision in a future work.

Finally, we briefly mention the angular dependence of the outflow.
The outflow is ejected approximately isotropically. 
However, the amount of the total mass and energy of the outflow are small near the rotational axis and the equatorial plane. 
This is due to the fact that the centrifugal force barrier and the torus prevent the outflow from propagating to the rotational axis and the equatorial plane, respectively.

We find that these properties of the outflow do not depend on $D_{\rm T}$. However, the total mass and energy of the outflow depend on $D_{\rm T}$. 
We analyze the dependence of $M_{\rm oflw}$ and $E_{\rm oflw}$ on $D_{\rm T}$  for models A2 and A4 in Appendix \ref{initp} 
(see also Table~\ref{t-o} for the result of this analysis). 

We conclude that the total energy of the outflow is $10^{54-56}$ erg and  approximately dominated by the radial component of the total kinetic energy, and the average velocity of the outflow is $\approx 0.2~c$ irrespective of initial chemical composition and rotation of the SMS core.

\section{discussion}
\label{discussion}
\begin{figure}[htbp]
	\includegraphics[width=\linewidth]{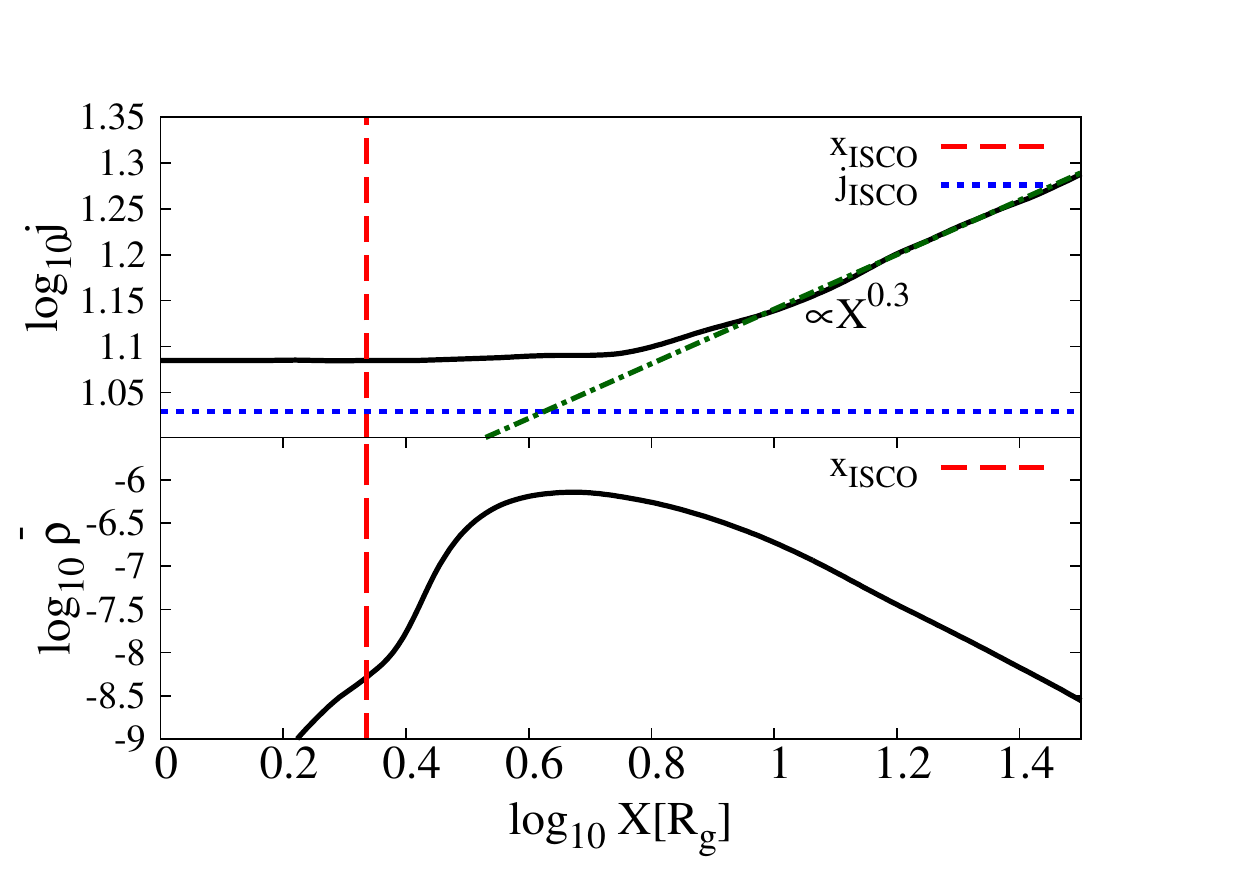}
	\caption{The profiles of the specific angular momentum (upper panel) and dimensionless density (lower panel) distribution as functions of the radial coordinate on the equatorial plane at $t-t_{\rm BH}= 9000t_{\rm g}$. The red-dashed and blue-dotted curves denote the radius of the ISCO and the specific angular momentum which is needed for a test particle to rotate at the ISCO, respectively. The green dashed-dotted curve denotes a slope proportional to $X^{0.3}$.}
	\label{fig:diskjr}
\end{figure}
A massive torus surrounding a black hole may be unstable to the so-called Papaloizou-Pringle instability~(PPI)~\cite{1984MNRAS.208..721P}. 
The PPI sets in if the torus has particular angular momentum distribution. 
If the PPI occurs, the torus is deformed by non-axisymmetric perturbation and gravitational waves (GWs) are emitted. 
Figure \ref{fig:diskjr} displays the density and specific angular momentum distribution of the torus along the cylindrical coordinate on the equatorial plane
for the N model with $\Gamma=1.335$ and $T_{\rm rot}/|W|\approx0.009$.
Immediately after the formation of the torus, the torus oscillates, and thus, we plot the profiles at $t-t_{\rm BH}=9000t_{\rm g}$, which is time  at which the oscillating perturbation of the torus is sufficiently damped. 
Here, $j$, $\bar{\rho}$, and $R_{\rm g}$ are the specific angular momentum, dimensionless density, and the gravitational radius of the black hole; $\bar{\rho}$ and $R_{\rm g}$ are defined by
\begin{eqnarray}
\bar{\rho} &\equiv& \rho K^N c^{-2N},
\end{eqnarray}
\begin{equation}
R_{\rm g} \equiv \frac{GM_{\rm BH}}{c^2},
\end{equation}
respectively. Here $K$ is the polytropic constant defined by Eq.~(\ref{s-14}). The peak of density depends weakly on $\Gamma$ and $T_{\rm rot}/|W|$ and is located at $X \approx 5R_{\rm g}$.
For $X \lesssim 5R_{\rm g}$, $j\approx {\rm const}$, and for $X \gtrsim 5R_{\rm g}$, $j \propto X^{0.3}$. 
The reason why $j\approx {\rm const}$ at $X \lesssim X_{\rm ISCO}$ is that a fraction of the torus material falls to the black hole 
from the inner edge of the torus due to the oscillation of the torus while conserving its specific angular momentum.

Previous general relativistic simulations show that the stability against the PPI depends on the profile of $j$ and strength of the self-gravity of tori~\cite{PhysRevD.83.043007}.
If a torus is strongly self gravitating and has the specific angular momentum distribution of the form of $j \propto X^\beta$, the torus is unstable to the PPI if $\beta \lesssim 0.25$ \cite{2011PhRvL.106y1102K}.
The torus found in our present study is also self gravitating and $\beta \approx 0$ at the density maximum, and hence, it may be unstable. 
To determine the stability to the PPI of these tori, we plan to perform a 3D numerical simulation in the future work. 

Here, we estimate the frequency and amplitude of GWs from the black hole-torus system assuming that the PPI hypothetically occurs. 
The amplitude of GWs can be approximately estimated by using quadrupole formula as
\begin{equation}
h_{ij} \approx \frac{2G}{c^4 r_{\rm L}} \ddot{Q}_{ij},
\label{quad}
\end{equation}
where $r_{\rm L}$ and $Q_{ij}$ are the luminosity distance and the mass quadrupole moment, respectively \cite{1973grav.book.....M}. 
We assume that a fraction of the torus deforms into one clump and rotates with Keplerian motion around the black hole. 
We regard this system as
a star with mass $M_{\rm clump}$ rotating around the central black hole with mass $M_{\rm BH} ( \gg M_{\rm clump})$. 
Then, the frequency of GWs measured by the observer, $f^{({\rm obs})}_{\rm gw}$, can be approximated by
\begin{equation}
f^{({\rm obs})}_{\rm gw}=\frac{1}{\pi (1+z)} \sqrt{\frac{GM_{\rm BH}}{r_{\rm tor}^3}},
\label{kep}
\end{equation}
where $z$ and $r_{\rm tor}$ are the cosmological redshift and the distance of the clump from the central black hole, respectively. 
In this system, the magnitude of $\ddot{Q}_{ij} $ can be approximated by $M_{\rm clump} v^2$, where $v$ is the orbital velocity of the clump. 
Previous numerical simulation shows that if the PPI occurs, the system emits quasiperiodic GWs and its emission will continue for $\gtrsim 100$ cycles 
if the viscosity of the torus is not very high \cite{2011PhRvL.106y1102K}. 
Hence the peak amplitude of GWs, $h_{\rm peak}$, may be enhanced by approximately $\sqrt{100}=10$ times.
Then if a SMS is located at $z=5$, we get  the typical peak frequency and amplitude of GWs
\begin{equation}
f^{({\rm obs})}_{\rm gw} \sim 1\times 10^{-2} \left (   \frac{z+1   }{6   }  \right )^{-1}\left (   \frac{ r_{\rm tor}  }{5R_{\rm g}   }  \right )^{-\frac{3}{2}}\left (   \frac{M_{\rm BH}   }{  10^5M_\odot }  \right )^{-1}{\rm Hz},
\label{d-2}
\end{equation}
\begin{eqnarray}
h_{\rm peak} \sim 2 \times 10^{-21} \left (   \frac{ M_{\rm tor}  }{0.05M_{\rm BH}   }  \right )\left (   \frac{ M_{\rm clump}  }{0.1M_{\rm tor}   }  \right )\left (   \frac{ M_{\rm BH}  }{ 10^5M_\odot   }  \right ) \nonumber \\
\times \left (   \frac{ r_{\rm tor}  }{5R_{\rm g}   }  \right )^{-1} \left (   \frac{  r_{\rm L} }{50{\rm Gpc}   }  \right )^{-1} \left (   \frac{N_{\rm cycle}   }{  100 }  \right )^{\frac{1}{2}}. \nonumber \\
\label{d-3}
\end{eqnarray}
Here, $M_{\rm tor}$ and $N_{\rm cycle}$ are the mass of the torus and the number of the cycles of GWs. 
From Eq.~(\ref{d-2}) and (\ref{d-3}), the typical frequency and amplitude of GWs are $\approx 10^{-2}$Hz and $\approx10^{-21}$. 
These values correspond to the sensitive observation band of the Laser Interfermometer Space Antenna~(LISA) \cite{2017arXiv170200786A} if a SMS is located at $z\leq 5$.

\section{conclusion}
\label{conclusion}
We explored the gravitational collapse of rotating SMS cores employing realistic initial conditions and including the effects of nuclear burning. 
We showed that though the efficiency of nuclear burning exponentially grows just before the black-hole formation, its effect is negligible. 
After the collapse, a fraction of the initial mass forms a torus surrounding the remnant black hole and drives an outflow. 
We also find that nuclear burning gives only a minor effect for the evolution of the torus.

We also studied the gravitational collapse of SMS cores with various adiabatic constants and rotation parameters without nuclear burning and analyzed quantitatively the mass of the torus, black-hole spin, and the property of the outflow. 
We found that the black-hole spin agrees well　with our previous analytical predictions \cite{0004-637X-818-2-157}. On the other hand, the mass of the torus is slightly smaller than the predicted value. 
This is because in our prediction, we regard each fluid element of the SMS core as a test particle, and assume that it has a circular orbit around the hypothetically formed black hole.
However, in reality, each fluid element of the SMS core feels the pressure force, and falls in an elliptical orbit to the central black hole, and hence, a fraction of the fluid element falls into the black hole even though they have specific angular momentum larger than the value of ISCO.

If a SMS core is sufficiently rapidly rotating ($T_{\rm rot}/|W|>0.004$), the outflow would have mass $\approx 1\%$ of initial mass and kinetic energy $10^{54-56}$ erg with its velocity $\approx 0.2c$. 
Exploring its time evolution and possibility for observing this outflow is our future work.

We also estimated the frequency and amplitude of GWs assuming that the torus surrounding the central black hole may be deformed by the PPI.
We found that if the PPI occurs, GWs emitted from the torus may be observed by LISA \cite{2017arXiv170200786A} if a SMS is located at $z \leq 5$.

{\em Acknowledgments}
:We are grateful to Koh Takahashi whose comments and suggestions about nuclear burning 
were of inestimable value for our study. 
We also owe a very important debt to Kenta Kiuchi who gave us invaluable comments about numerical calculation. 
We also thank Yudai Suwa for a helplful discussion. 
Numerical computations were performed on the supercomputer XC30 at CfCA of NAOJ, and XC40 at YITP of Kyoto University. 
This work was supported by Grant-in-Aid for Scientific Research~(Grants No.~16H02183~16K17706~16H05341~15H00782) of Japanese MEXT/JSPS. 

\appendix
\section{Calculation of the nuclear burning}
\label{app1}
The effect of nuclear burning might be able to be taken into account by 
naively adding the nuclear energy generation rates as a source term of the evolution equation of the fluid with fixed chemical composition as~\cite{2012ApJ...749...37M}
\begin{eqnarray}
\nabla_\mu T^{\mu \nu}&=& \frac{\rho_{0} \dot{q}u^\nu}{c}, \label{app-1}\\
\dot{q}&=&\dot{q}_{\rm CNO}+ \dot{q}_{3\alpha}. \label{app-2}
\end{eqnarray}
However, we find that this method is not very appropriate to calculate the effects of nuclear burning even if the change of the chemical composition is negligible.

To illustrate this fact, we compare the results of two simulations with the same initial conditions but with different formalisms to incorporate the effects of nuclear burning. 
One is the formalism described in Eq.~(\ref{app-1}) (model T1), and the other is that described in Sec.~\ref{nuc} (model T2).
We set the initial condition which mimics model R1.d of Ref.~\cite{2012ApJ...749...37M}, that is, the SMS core is rotating at mass-shedding limit, its mass is $\approx 5\times 10^5M_\odot$, and its chemical composition is $(Y_{\rm p},~X_{\alpha},~ Z_{\rm C}) = (0.748,~0.250,~0.002)$. 
We performed another simulation with the same condition as model T2 except initial metallicity chosen as $(Y_{\rm p},~X_{\alpha},~ Z_{\rm C}) = (0.74,~0.25,~0.01)$ (model T3).
\begin{figure}[htbp]
	\includegraphics[width=\linewidth]{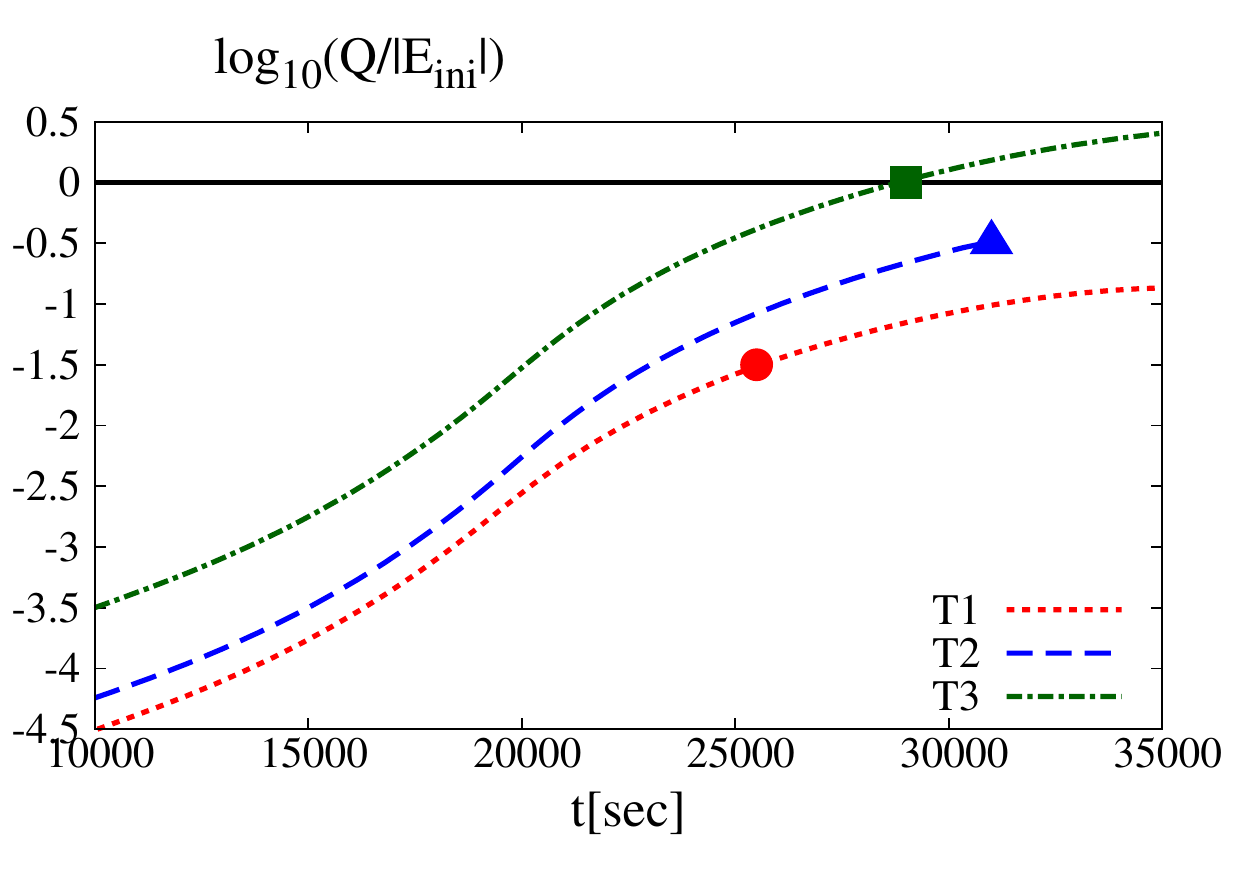}
	\caption{Time evolution of the ratio of the released rest-mass energy 
	due to the nuclear burning to the initial total energy. The dotted, dashed, and dashed-dotted curves denote for models T1, T2, and T3, respectively. For models T1 and T3, the explosion is found at $t \approx 25000$ s (filled circle) and $t\approx 29000$ s (filled square), while for model T2, a black hole is formed at $t\approx 31000$ s (filled triangle), respectively.} 
	\label{fig:appen}
\end{figure}

Figure~\ref{fig:appen} shows the time evolution of the ratio of the released rest-mass energy due to the nuclear burning, $Q$, to the initial total energy, $E_{\rm ini}$. 
Here, $Q$ and $E_{\rm ini}$ are defined by
\begin{eqnarray}
Q(t)\equiv \left\{ \begin{array}{ll}
\displaystyle \int_{0}^{t} dt \int_V \rho_{0*} \dot{q} dV & \text{(T1)}, \\
\displaystyle \text{Eq.~(\ref{s-21.2})} & \text{(T2 and T3)}, \\
\end{array} \right.
\end{eqnarray}
and
\begin{equation}
E_{\rm ini} \equiv (T_{\rm tot}+U+W) (t=0~{\rm s}),
\label{a-4}
\end{equation} 
respectively.
The filled circle denotes the time ($t\approx 25000 $ s) at which the system becomes unbound and starts exploding for model T1.
We check that in the absence of the nuclear burning, this model does not explode. 
Thus this explosion is due to the nuclear burning. 
However, the value of $Q$ is by more than one order of magnitude smaller than the value of $|E_{\rm ini}|$ at this time. 
Thus, in terms of energy conservation, it is incomprehensible that the nuclear burning causes the explosion.
Indeed, we find that for model T2, the nuclear burning cannot halt the collapse and the black hole is formed at $t\approx 31000$ s (filled triangle).
We check that the chemical composition at the center of the SMS core for model T2 does not change approximately until $t\approx25000$ s. 
Thus this difference would be due to the fact that Eq.~(\ref{app-1}) violates the energy-momentum conservation. 
We suspect that even though the violation is much smaller than the total energy, this violation is  accumulated and finally induces 
the unnatural increase of the total energy.

We also find that for model T3, the SMS core starts exploding at $t\approx 29000$ s (filled square).
At this time, $Q \approx |E_{\rm ini}|$ is realized, and thus, this explosion would be indeed due to the effect of nuclear burning. 
We conclude that for the computation of the gravitational collapse of a SMS core, 
it is important to employ the formalism in which the conservation of the total energy is satisfied with the required level. 

\section{Definition of the mass and energy flux of the outflow}
\label{app2}
In this section, we define the mass and energy fluxes of the outflow by using the conservation of the total mass and energy.
Local conservation equaion for the energy-momentum tensor together with the Kililng equation for a timelike Kililng vector field $\xi^\mu$  give
\begin{equation}
\nabla_\mu (T^\mu_\nu \xi^\nu)=0.
\end{equation}
For the analysis of the outflow, we assume that the space-time becomes a stationary state after the collapse. 
Then $(\partial_t)^\mu$ should be a timelike Killing vector field, and hence,
\begin{equation}
\nabla_\mu T^\mu_t=0,
\label{kilt}
\end{equation}
is realized.
On the other hand, if we neglect the nuclear burning after the collapse, the rest-mass density should be conserved, i.e.,
\begin{equation}
\nabla_\mu (\rho_0 u^\mu)=0.
\label{kilr}
\end{equation}

By using Eqs.~(\ref{kilt}) and (\ref{kilr}), we get two equations of continuity as 
\begin{equation}
\frac{\partial{  \rho_{ A} } }{\partial{  t  } }  + {\bm \nabla} \cdot {\bm f}_{\rm A}=0 \ \ ({ A=M, E}),
\label{eoc}
\end{equation}
where
\begin{eqnarray}
\rho_{ M} &\equiv& \rho_{0*}, \\
\rho_{ E} &\equiv& -\rho_{0*}c^2 \left (  \frac{h_0 u_t}{c^3} + \frac{ P  }{ \rho_0c^3 u^t  }+1  \right ),
\end{eqnarray}
and 
\begin{eqnarray}
{\bm f}_{ M} &\equiv& \rho_{0*} {\bm v}, \\
{\bm f}_{ E} &\equiv& -\rho_{0*}c^2 {\bm v}\left ( \frac{ h_0 u_t}{c^3} +1 \right).
\end{eqnarray}
Here, $\rho_{0*} \equiv \rho_0 \sqrt{-g} c u^t$ and $h_0$ are the weighted rest-mass density and the specific enthalpy, respectively. 
$\rho_{\rm E}$ is defined as the total energy density minus rest-mass energy density. 
Subscripts "$M$" and "$E$" denote the total mass and energy, respectively. 

Assuming that the internal energy is much smaller than the kinetic and potential energy of the outflow, we define the fluid component  which satisfies $u_t < -1$ as the outflow component.
In Newtonian approximation, $u_t$ is approximated by
\begin{equation}
u_t \approx -1 + \frac{GM_{\rm BH}}{R}- \frac{1}{2} v^2,
\end{equation} 
where $M_{\rm BH},~R$, and $v$ are the mass of the black hole, distance from the black hole, and velocity of the fluid element, respectively. Hence $u_t < -1$ is equivalent to $1/2v^2 > GM_{\rm BH}/R$ which implies that the fluid element is unbound.

To analyze the mass and energy of the outflow, we replace $\rho_0$ with $\rho_0^{\rm eje}$ where $\rho_0^{\rm eje}$ is defined by
\begin{equation}
\rho_0^{\rm eje} \equiv \rho_0 \Theta (-u_t-1),
\end{equation}
\begin{eqnarray}
\Theta(x)\equiv \left\{ \begin{array}{ll}
\displaystyle 1 & x\geq 0, \\
\displaystyle 0 & x<0. \\
\end{array} \right.
\end{eqnarray}
We denote the quantities associated with the unbound material by subscript "eje".

Then, the mass and energy of the outflow, which are ejected from the domain $\sqrt{X^2+Z^2}<D$, can be written as
\begin{eqnarray}
A_{\rm eje} (D,t) &=& -\int_0^t dt' \int {\bm f}_{ A}^{\rm eje} \cdot d {\bm S} \label{ejected} \\ 
&=& \int_0^t dt' \int_0^\frac{\pi}{2} d\theta (-4\pi D^2 {\bm f}_{ A}^{\rm eje}\cdot \hat{r}{\rm sin}\theta) \ \ ({ A=M, E}), \nonumber
\end{eqnarray}
where $\hat{r}$ and $\theta$ are the radial unit vector and the polar angle, i.e., ${\rm tan}\theta= X/Z$. 
We define the total mass and energy of the outflow by
\begin{equation}
A_{\rm eje} (D)\equiv A_{\rm eje}~(D,t=t^{\rm *}),
\label{tot}
\end{equation}
where $t^*$ is the time at which all of the unbound fluid finishes escaping from the domain $\sqrt{X^2+Z^2}< D$.

The total mass and energy of the outflow also can be defined by
\begin{equation}
A_{\rm oflw}(t) \equiv \int_{V'} \rho_A^{\rm eje} dV~(A=M,E).
\label{total}
\end{equation}
$A_{\rm oflw}(t=\infty)$ should correspond to $A_{\rm eje}(D)$ irrespective of the value of $D$ if the total mass and energy of the outflow is conserved. 
Finally, we define the total kinetic energy, $T_{\rm oflw}$, the radial component of the total kinetic energy, $T_{\rm oflw}^r$, the total internal energy, $U_{\rm oflw}$, and the total gravitational potential energy, $W_{\rm oflw}$, of the outflow by
\begin{eqnarray}
T_{\rm oflw}(t) &\equiv& \int_{V'} \frac{1}{2} \rho_{0*}^{\rm eje} c^2 (1-(\alpha c u^t)^2) dV, \label{ttot}\\
T_{\rm oflw}^r(t) &\equiv& \int_{V'} \frac{1}{2c} \rho_{0*}^{\rm eje} h_0 v^r u_r dV, \label{trtot}\\
U_{\rm oflw}(t)&\equiv& \int_{V'} \rho_{*}^{\rm eje} \epsilon dV, \label{utot}
\end{eqnarray}
and
\begin{equation}
W_{\rm oflw}(t)\equiv E_{\rm oflw} - T_{\rm oflw}- U_{\rm oflw},
\label{wtot}
\end{equation}
respectively. Here, $\rho^{\rm eje}_* \equiv \rho \sqrt{-g} c u^t \Theta(-u_t-1)$.

\section{Initial perturbation dependence of the outflow}
\label{initp}
In this section, we analyze the dependence of total mass and energy of the outflow on $D_{\rm T}$. 
To see this, we performed additional simulations with $T_{\rm rot}/|W|=0.005$ for the helium-burning models (model A5). 

\begin{figure}[htbp]
	\includegraphics[width=\linewidth]{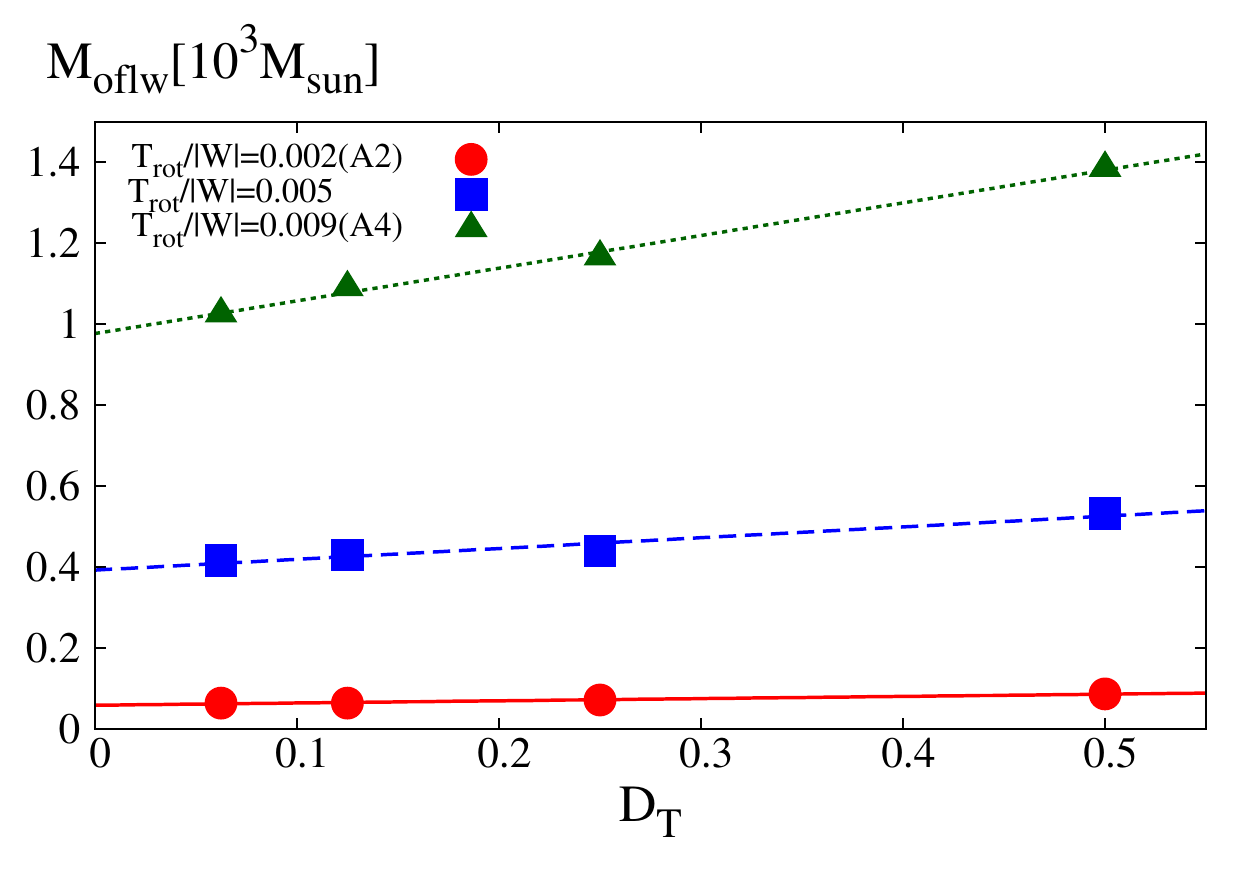}
	\caption{The total mass of the outflow, $M_{\rm oflw}$, as a function of $D_{\rm T}$ with $T_{\rm rot}/|W|= 0.002$ (model A2, filled-circles), $0.005$ (model A5, filled-squares), and $0.009$ (model A4, filled-triangles) at $t-t_{\rm BH}=5000t_{\rm g}$ for the helium-burning models. The solid, dashed, and dotted curves denote the linear fitting function for the value of  $T_{\rm rot}/|W|=0.002$, $0.005$, and $0.009$, respectively.}
	\label{fig:mot}
\end{figure}

\begin{figure}[htbp]
	\includegraphics[width=\linewidth]{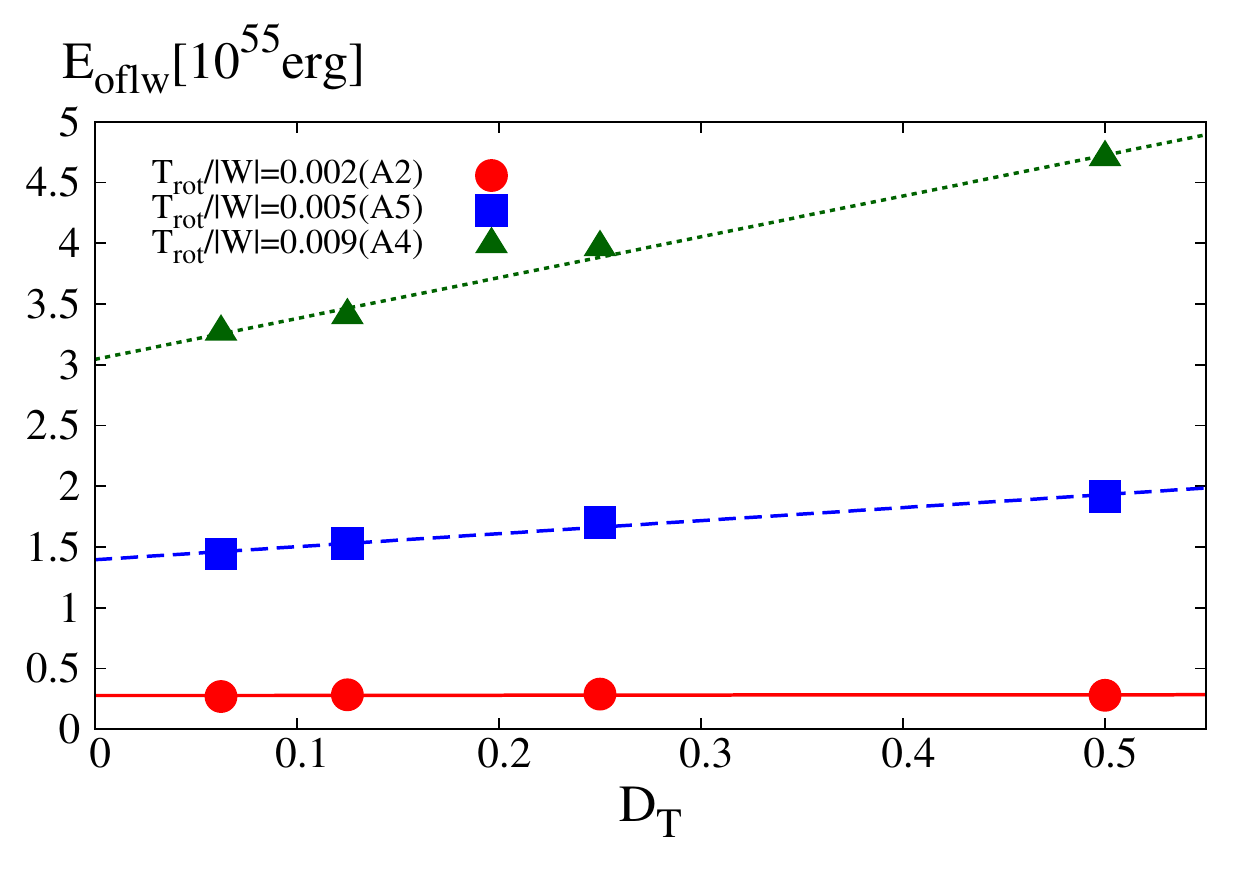}
	\caption{The total energy of the outflow, $E_{\rm oflw}$, as a function of $D_{\rm T}$ with $T_{\rm rot}/|W|= 0.002$ (model A2, filled-circles), $0.005$ (model A5, filled-squares), and $0.009$ (model A4, filled-triangles) at $t-t_{\rm BH}=5000t_{\rm g}$ for the helium-burning models. The solid, dashed, and dotted curves denote the linear fitting function for the value of  $T_{\rm rot}/|W|=0.002$, $0.005$, and $0.009$, respectively.}
	\label{fig:tot}
\end{figure}

Figures \ref{fig:mot} and \ref{fig:tot} display the dependence of the total mass and energy of the outflow  on $D_{\rm T}$ for the helium-burning models with three rotation parameters, i.e., $T_{\rm rot}/|W|=0.002~({\rm model~A2}), 0.005~({\rm model~A5})$, and $\approx 0.009~({\rm model~A4})$ at $t- t_{\rm BH}=5000t_{\rm g}$, respectively. 
Here, $M_{\rm oflw}$ and $E_{\rm oflw}$ are defined by Eq.~(\ref{total}).
The total mass and energy of the outflow increases approximately linearly with $D_{\rm T}$. 
Thus we can estimate the total mass and energy of the outflow of $D_{\rm T}=0$ by using linear fitting formula. 
Each of three lines of Figs.~\ref{fig:mot} and \ref{fig:tot} denote the linear fitting function for each value of $T_{\rm rot}/|W|$. 
Here the value with $D_{\rm T}=0$ should be regarded as the physical value.

We find that for $T_{\rm rot}/|W| =0.002$ (model A2) , $0.005$ (model A5), and $\approx 0.009$ (model A4), the value of $M_{\rm oflw}$  is overestimated with $D_{\rm T}=0.5$  by $47\%$, $35\%$, and $42\%$, and $E_{\rm oflw}$ is overestimated by $1\%$, $37\%$, and $54\%$, respectively. 

We speculate the reason why $M_{\rm oflw}$ and $E_{\rm oflw}$ are increasing functions of $D_{\rm T}$ is as follows: For the larger value of $D_{\rm T}$, 
the fluid falls with larger velocity, and hence, shock heating more efficiently occurs around the inner edge of the torus, and the bubble which we describe in Sec.~\ref{formation} is more strongly pushed by the torus.

\end{document}